\documentclass[a4paper]{article}
\usepackage{graphicx}
\usepackage{amsmath,amssymb}

\newcommand\beq{\begin{equation}}
\newcommand\eeq{\end{equation}}
\newcommand\bea{\begin{eqnarray}}
\newcommand\eea{\end{eqnarray}}
\begin{document}

\vspace{-2.0cm}
\bigskip

\centerline{\Large \bf  Loop Approach to Lattice Gauge Theories} 
\vskip .8 true cm

\begin{center} 
{\bf Manu Mathur}\footnote{E. Mail: manu@bose.res.in}  

S. N. Bose National Centre for Basic Sciences \\
 JD Block, Sector III, Salt Lake City,  Calcutta 98, India.

\end{center} 
\bigskip

\centerline{\bf Abstract}

We solve the Gauss law and the corresponding Mandelstam 
constraints in the loop Hilbert space ${\cal H}^{L}$ 
using the prepotential formulation of $(d+1)$ dimensional SU(2) 
lattice gauge theory. The resulting orthonormal and complete 
loop basis, explicitly constructed in terms of the $d(2d-1)$ prepotential 
intertwining operators, is used to transcribe the gauge dynamics 
directly in ${\cal H}^{L}$ without any redundant gauge and loop degrees of 
freedom. Using generalized Wigner-Eckart theorem and Biedenharn
-Elliot identity in ${\cal H}^L$, we show that the above loop dynamics for 
pure SU(2) lattice gauge theory  in arbitrary dimension, is given by  
real and symmetric $3nj$ coefficients of the second kind (e.g., n=6, 10 for 
d=2, 3 respectively). 
The corresponding ``ribbon diagrams" representing SU(2) loop dynamics  
are constructed. The prepotential techniques are trivially extended to 
include fundamental matter fields leading to a description in terms of loops 
and strings. The SU(N) gauge group is briefly discussed. 

\section{\bf Introduction}

The idea that  gauge theories should be formulated completely in terms of  loops in space 
carrying electric fluxes is quite old, appealing and has long history 
\cite{mans1,wilson,yang,rest,kogut,giles}.  In the context of electrodynamics,  Yang \cite{yang} 
 has emphasized the importance of  path dependent ``non-integrable phase factors" 
carrying electric fluxes to provide a complete description of all the quantum effects.  
In the context of quantum chromodynamics the loop formulation, 
without any colored gluon or colored quark degrees of freedom, is further expected to 
provide a better framework to analyze non-perturbative low energy issues like 
color confinement.
In the context of gravity, the relatively recent Hamiltonian formulation 
of quantum gravity in terms of SU(2) connections has also been reformulated 
in terms of loops leading to loop quantum gravity \cite{rovbook}. 
Therefore, the loop formulation of gauge theories may eventually 
provide a common geometrical platform to understand all interactions in 
nature.  In this work, we analyze some of the basic kinematical as well as dynamical issues 
involved in the Hamiltonian formulation of lattice gauge theories in terms of loops. 

The standard construction of the loop states for pure SU(N) gauge theory on lattice 
\cite{kogut} consists of considering the set of all  oriented loops  $\Gamma$ 
and constructing the corresponding gauge invariant Wilson loop operators 
$W_{{\gamma}} \equiv {\textrm Tr}U_{{\gamma}}$ for every ${\gamma} 
\in \Gamma$. Acting on the strong 
coupling vacuum, all possible gauge invariant operators of the form 
$W_{{\gamma}_1}W_{{\gamma}_2}....W_{{\gamma}_m}$ create all possible gauge invariant 
states associated with the corresponding loops ${\gamma}_1{\gamma}_2...{\gamma}_m$.  
These loop states are manifestly gauge invariant, geometrical and form a basis, usually 
known as Wilson loop basis (see section 2.1). However, a  serious problem with this Wilson 
loop basis is that it again over-describes gauge theory.
This time, the over description is because loops of all shapes and sizes have to 
be included in constructing the above loop basis. Therefore, 
one is again confronted with too many redundant (but now SU(N) gauge invariant) 
loop degrees of freedom (see section 2.1).  
The Mandelstam constraints \cite{mans1} amongst the various 
loop states express this over-completeness of the Wilson loop basis 
(see section 2.1).  As these constraints represent the linear 
dependence of the gauge invariant states  associated with loops of all sizes,
they are highly non-local and hence difficult to solve (see section 2.1). 
In the strong coupling ($g \rightarrow \infty$) limit, the 
loops are small and they carry small electric fluxes \cite{kogut}. Therefore, 
the Mandelstam constraints can be easily solved by using Gram-Schmidt 
orthogonalization procedure amongst the small number of loop states considered
within this ($g \rightarrow \infty$)  limit. 
However, in the continuum ($g \rightarrow 0$) limit, as opposed to the strong 
coupling limit, large loops carrying large electric fluxes will be important \cite{furma}. 
Therefore, the problem of over-completeness of the Wilson loop basis  
will become more and more acute as we remove the lattice cut-off and approach the 
continuum limit. This over-completeness, in turn, will result in rapid proliferation 
of spurious zero modes of the Hamiltonian in the Wilson loop basis. 
Therefore, the initial problem in loop formulation 
is to solve the Mandelstam constraints {\it exactly} before addressing any dynamical 
issue.  Infact, as stated by Gambini and Pullin in \cite{gamb}\footnote{Chapter 12, page 
303-304}: ``{\it The proliferation of loops when one considers larger lattices and higher 
dimensions completely washes out the advantages provided by the (loop) formalism.}". 

\noindent The motivation and purpose of the present work is to systematically develop 
ideas and techniques to reformulate lattice gauge theories in loop space without any 
spurious loop degrees of freedom. We solve SU(2) Mandelstam constraints leading to an 
orthonormal loop basis which is complete and characterized exactly by $3(d-1)$ angular 
momentum quantum numbers per lattice site \cite{manu2}. Further, we show that the most 
economical description of the pure $(d+1)$ SU(2) lattice gauge theory dynamics, involving 
only the relevant loop degrees of freedom, is given by $3nj$ coefficients of the second 
kind and therefore it is highly geometrical. 

\noindent We will work within the prepotential formulation \cite{manu1,manu2}
of SU(2) lattice gauge theory Hamiltonian \cite{kogut}. 
The prepotential approach has extended $SU(2) \otimes U(1)$ gauge invariance. 
The prepotential operators are SU(2) harmonic oscillator doublets attached to 
the initial and the final points of every link. They are connected by  
the U(1) gauge invariance mentioned above (see section 2.2). The advantage 
of prepotential approach is that the non-local and over-complete Wilson loop 
basis can be constructed and studied locally in terms of the SU(2) invariant 
$d(2d-1)$ ``prepotential intertwining operators" at every lattice site. This local 
description of the Wilson loop basis is characterized by  $2d(d-1)$  (integer) 
intertwining quantum number per lattice site. Further, the Mandelstam constraints, 
which appear highly non-local in terms of 
the link operators, become local in terms of the prepotential  
operators (see section 2.3.1).  This enables us to solve them explicitly 
using simple group theoretical ideas \cite{sharat1,sharat2}. The novel U(1) gauge 
invariance of the prepotential formulation then 
enables us to explicitly construct  an {\it orthonormal and complete loop basis}  
in the entire loop Hilbert space ${\cal H}^{L}$ on the lattice. 
As expected, this orthonormal basis is characterized by $3(d-1)$ 
(as opposed to $2d(d-1)$ quantum numbers for 
the Wilson loop basis) gauge invariant  quantum numbers per lattice site.  
Having solved all the constraints, we use the generalized Wigner-Eckart theorem 
and Biedenharn-Elliot identities in the resulting orthonormal loop basis to compute the 
loop dynamics. Our approach also enables us to compute loop dynamics locally site by site. 
The final results at different lattice site are then glued together through the U(1) gauge 
invariance. 
We show that in $d+1$ dimension the SU(2) loop dynamics in the 
$(I,J)$ plane where $I,J=1......d$ and $I< J$ plane is given by $3nj$ 
coefficients of the second kind where $n=2\left[2+d +(J-I) -\delta_{I,1}
-\delta_{J,d}\right]$.  The matrix elements of the Hamiltonian 
in the above loop basis are found to be real and symmetric.  
The $3nj$ coefficients of second kind, and therefore the loop dynamics, are 
graphically represented by the ribbon diagrams. 

The plan of the paper is as follows.  In the first half (section 2) 
we discuss the kinematical issues and in the second half 
(section 3) we discuss the dynamical issues. In both these sections, 
the explicit computations are done in $d=2$. This keeps the discussions simple 
and also illustrates all the essential ideas and techniques involving prepotentials. 
Their generalization to  arbitrary d dimension is obvious and done next. 
As the complications caused by over-completeness of the loop basis or equivalently 
the Mandelstam constraints have been major obstacles in the loop approach to gauge theories, 
we first review them on lattice in section 2.1.  
In section 2.2, we  briefly discuss the $SU(2) \otimes U(1)$ gauge invariant 
prepotential approach \cite{manu1} which enables us to cast the Mandelstam 
constraints in a simple local form. 
In section 2.3, we solve the Mandelstam constraints and give 
all possible orthonormal loop  state solutions \cite{manu2} in terms of the $d(2d-1)$ 
prepotential intertwining operators. In section 2.4, we discuss 
inclusion of matter fields leading to a gauge theory description in terms of loops and 
strings.  In section 3.1 and 3.2, we  compute the matrix elements of the Hamiltonian in the 
above loop basis and discuss the ribbon diagrams representing these amplitudes.     
In section 4, SU(N) gauge group is briefly discussed. 
The techniques used in constructing the orthonormal loop 
state basis are given in appendix A. The technical details involved in computing  
loop dynamics are given in appendix B. 
  
\section{The kinematical issues}  

This section is devoted to the kinematical issues involving the loop states in 
pure SU(2) lattice gauge theory. We first discuss the Mandelstam 
constraints in terms of the original lattice link operators. They look highly non-local. 
We then cast them in their local form at every lattice site n in terms of the SU(2) invariant 
prepotential intertwining operators at n.  Next, we convert the problem of solving these 
{\it local} Mandelstam constraints to the problem of finding common (orthonormal) eigenvectors 
of a complete set of commuting observables (C.S.C.O) containing $(4d - 3)$ angular momentum 
operators at n. In particular, we show that the states related by the Mandelstam constraints 
at  n are degenerate with respect to a subset of operators belonging to the above C.S.C.O.. 
Therefore, the common eigenvectors of the C.S.C.O at n, which are specific 
linear combinations of the states related by the Mandelstam constraints, 
lead us to an orthonormal local basis at n.  The final orthonormal loop basis over the 
entire lattice, characterized by  $3(d-1)$ angular momentum quantum
numbers per lattice site, is obtained by weaving or gluing these local orthonormal 
basis at different lattice sites according to the additional  U(1) Gauss law    
associated with the prepotential formulation. 
 
\subsection{Mandelstam constraints on lattice} 

\noindent 
On lattice the number of  gauge invariant degrees of freedom 
$({\cal N})$ is given by the dimension of the quotient space 
$\otimes_{links} SU(2) / \otimes_{sites} SU(2)$. 
Thus for a d-dimensional 
periodic lattice with $n^{d}$ sites and $dn^{d}$ links: 
\bea 
{\cal N} = 3(d-1)n^{d}.    
\label{ndf}
\eea
Therefore, a complete description of the SU(2) gauge invariant physical 
Hilbert space on a periodic lattice should require ${\cal N}$ quantum numbers 
or equivalently $3(d-1)$ quantum numbers per lattice site. 
However, the Wilson loop basis is characterized by 
${\cal M} = 2d(d-1)n^{d}$ linking or intertwining quantum 
numbers (see section 2.3). Thus there are  ${\cal M} - {\cal N}  
= \left(2d^2 -5d +3\right) n^d$ redundant degrees of freedom.  
This clearly establishes that the  Wilson loop basis is over-complete.  
Further, the redundant degrees of freedom or degree of over-completeness 
increases as $d^2$ for large d.  Let us illustrate this over-completeness 
by giving the simplest example in $d=2$. We consider two plaquettes A and B  
touching each other at a common lattice site n as shown in Figure (\ref{fig1}a). 
The corresponding Wilson loop operators satisfy: 
\begin{figure}[b]
\begin{center}
\includegraphics[width=0.8\textwidth,height=0.25\textwidth]
{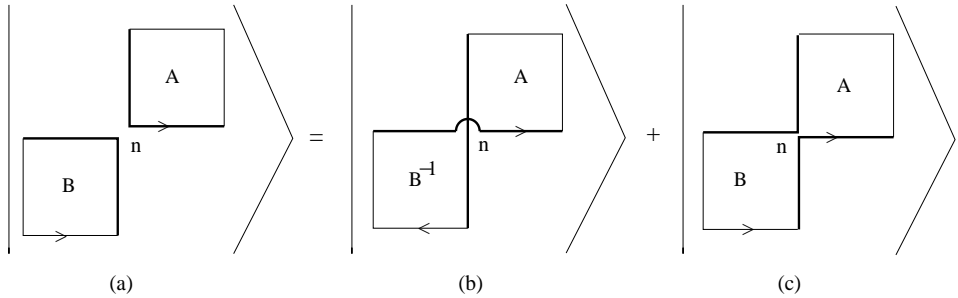}
\end{center}
\vspace{-5mm}
\caption{The graphical representation of the simplest Mandelstam constraint 
(\ref{3sr}) with $N_A = N_B =1$. The thick lines are for the 
later comparison of the same in the prepotential formulation.} 
\label{fig1}
\end{figure}
\bea 
\left(\textrm{Tr} U_{A}\right) \left(\textrm{Tr} U_{B}\right)  \equiv 
\textrm{Tr} \left(U_{A}U_{B}\right) + \textrm{Tr} \left(U_{A}U_{B}^{-1}\right).  
\label{mc} 
\eea
The relation (\ref{mc}) is a trivial identity involving any two SU(2) 
matrices $U_{A}$ and $U_{B}$. It can be checked by writing $U_{X} = 
X_{0} 1 + i \sum_{a=1}^{3} X_{a} \sigma^{a}$ where $\sigma^{a}$ 
are the Pauli matrices, $X_{0}, X_{a}$ are real and satisfy 
$X_{0}^2+X_{1}^2+X_{2}^2+X_{3}^2 =1$.  The generalisations of the 
identity (\ref{mc}) to the other classical Lie groups have
been discussed in \cite{gli}. We define the following three loop states:
\bea  
|\gamma_1\rangle  \equiv 
\left(\textrm{Tr} U_{A}\right)\left(\textrm{Tr} U_{B}\right)|0\rangle, ~ 
~~~~
|\gamma_2\rangle  \equiv 
\textrm{Tr} \left(U_{A} U_{B}^{-1} \right)|0\rangle,  
~~~~
|\gamma_3\rangle  \equiv \textrm{Tr} \left(U_{A} U_{B}\right)|0\rangle,  
\label{3s}
\eea
The identity (\ref{mc}) implies the simplest Mandelstam constraints: 
\bea 
|\gamma_1\rangle = |\gamma_2\rangle + |\gamma_3\rangle.  
\label{3sr}
\eea
Thus we see that the three loop states $|\gamma_1 \rangle, |\gamma_2 \rangle$ and 
$|\gamma_3 \rangle$ are linearly dependent.  
To appreciate the problem further, let us consider most general  loop states 
involving only these two plaquettes A and B: 
\bea 
|N_A,N_B\rangle & \equiv & (\textrm{Tr} U_{A})^{N_A} (\textrm{Tr}U_{B})^{N_B} 
|0\rangle  \nonumber \\
          & = &  (\textrm{Tr} U_{A})^{N_A-1} (\textrm{Tr} U_{B})^{N_B-1} 
(\textrm{Tr}U_{A}U_{B} + \textrm{Tr}U_{A}U_{B}^{-1}) |0\rangle  \nonumber \\  
          & = &  (\textrm{Tr}U_{A})^{N_A-2} (\textrm{Tr}U_{B})^{N_B-2} 
(\textrm{Tr} U_{A}U_{B} + \textrm{Tr} U_{A}U_{B}^{-1})^{2} |0\rangle \nonumber \\  
          &.& \nonumber \\ 
          &.& \nonumber \\ 
          & = &  (\textrm{Tr} U_{A})^{N_A-N_{min}} (\textrm{Tr} U_{B})^{N_B-N_{min}} 
(\textrm{Tr}U_{A}U_{B} + \textrm{Tr}U_{A}U_{B}^{-1})^{N_{min}} |0\rangle    
\label{mcc} 
\eea
where $N_{A}$, $N_{B}$ are two arbitrary integers representing the angular momentum 
fluxes over  A and B and $N_{\rm min}$=Minimum$(N_A,N_B)$. 
Thus, given the loop state $|N_A,N_B\rangle$, we have produced 
$\left(2N_{\rm min}+1\right)$ distinct but linearly dependent Wilson loop states 
contained in  the relations (\ref{mcc}).  
Therefore, even in $d=2$, for the simplest loop 
states over the two plaquettes, the number of linearly dependent loop states 
increases with the SU(2) flux value $N_{\rm min}$. Note that we have to include 
states with $N_{min}$ taking arbitrarily large values.  Next one can imagine
extending the picture in Figure (\ref{fig1}) by adding more plaquettes leading to
further constraints\footnote{In \cite{giles} it is shown that
the Mandelstam constraints constitute sufficient algebraic conditions on Wilson loop
variables to allow reconstruction of the corresponding  gauge potentials.}.
In general, {\it one has to address 
the problem of finding a complete set of linearly independent loop states amongst the 
loop states of all shapes, sizes carrying arbitrary fluxes and 
touching/crossing one another at arbitrary number of lattice sites}. Further, it is clear that 
this problem becomes more and more difficult as the dimension d increases.  
This is the reason why the Mandelstam constraints have been notorious and major 
obstacle in the development of the loop formulation of gauge theories \cite{gamb}.  
The solutions of the Mandelstam constraints have been discussed at various levels of 
approximations and difficulties in the past.  In \cite{gamb, gambini} an approximate 
loop cluster method in 2+1 dimensions is developed and the Schr\"odinger equation is 
written down as difference equations in these cluster coordinates. However, approximations 
involved are uncontrolled \cite{gamb} to enable us to go to 3+1 dimensions. 
Motivated by strong coupling expansions, various methods involving cluster 
of loops and their truncations to define Hamiltonian eigenvalue problem have been 
developed to go beyond strong coupling region \cite{rest2}. 
In \cite{brugmann} the Mandelstam constraints are solved and eigenvalues equations
are analyzed on computer using small lattices and small loops. 
In \cite{loll,watson} the Mandelstam constraints are solved classically  
on a finite periodic d=2, 3 lattice leading to an independent loop configuration space. 
However, the issues like quantization of these loop  variables and setting 
up the corresponding Schr\"odinger equation are not clear \cite{watson}. We now briefly 
review the various operators 
in the Kogut-Susskind formulation \cite{kogut} and define the prepotential operators 
\cite{manu1} which enable us to solve the Mandelstam constraints exactly and explicitly 
in arbitrary dimensions.   

\subsection{\bf The prepotential operators}

\begin{figure}[t]
\begin{center}
\includegraphics[width=0.9\textwidth,height=0.15\textwidth]
{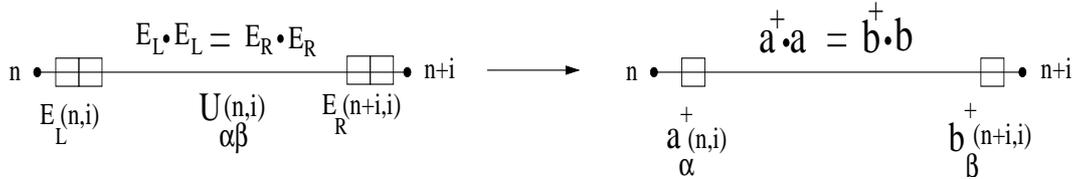} 
\end{center}
\vspace{-5mm}
\caption{(a) The original $\left(E_L^{a}(n,i), E_R^{a}(n+i,i), U_{\alpha\beta}(n,i)\right)$ 
operators, (b) the new $\left(a_{\alpha}(n,i),b_{\alpha}(n+i,i)\right)$ operators on the 
link $(n,i)$. The electric fields $E_{L}, E_{R}$ transform in the adjoint representation. 
The prepotentials 
$a_{\alpha}(n,i), b_{\alpha}(n+i,i)$ transform as SU(2) fundamental doublets at lattice site 
n and n+i respectively. Therefore, we represent them 
by Young tableau $\square$ at n and n+i respectively.}
\label{fig2}
\end{figure}

The kinematical variables involved in Kogut and Susskind Hamiltonian 
formulation \cite{kogut} of lattice gauge theories describe SU(2) rigid 
rotators attached to every link $(n,i)$ of the lattice. The kinematical 
variables are: 
a) SU(2) link operators $U(n,i) = 
\left( \begin{array}{ccc} 
U_{11}(n,i) & U_{12}(n,i)     \\
U_{21}(n,i)  & U_{22}(n,i)\\
\end{array} \right)$, describing the orientation of the body fixed frame of the 
rigid body from the space fixed frame, b) the electric fields 
$E^{a}_{L}(n,i)$ and $E^{a}_{R}(n+i,i)$ 
which  are the  components of the angular momentum  
in the body fixed and space fixed frames respectively. 
This description is shown in  Figure (\ref{fig2}a).   
The link operators $U(n,i)$ satisfy the SU(2) conditions:  
\bea 
U(n,i)U^{\dagger}(n,i) = U^{\dagger}(n,i)U(n,i) = {\cal I}, ~~  |U(n,i)| =1.  
\label{det1}
\eea
Above ${\cal I}$ is $2 \times 2$ identity matrix and $|U| \equiv {\rm det} U$. 
The rigid body commutation relations are\footnote{In \cite{manu1}, we had used 
$E^a(n,i)= -E^a_{L}(n,i)$ and $e^a(n,i) \equiv E^a_R(n+i,i)$ resulting in an extra -ve 
signs in the commutation relations involving $E^a_{L}$ in (\ref{ccr}).} \cite{kogut}:  
\bea 
\left[E^{a}_{L}(n,i),U(n,i)\right]  =   - \left(\frac{\sigma^a}{2}\right)~ U(n,i) & \Rightarrow & 
\left[E^{a}_{L}(n,i),E^{b}_{L}(n,i)\right]  =  i\epsilon^{abc} E^{c}_{L}(n,i), 
\nonumber \\
\left[E^{a}_{R}(n+i,i),U(n,i)\right]  =  U(n,i) \left(\frac{\sigma^a}{2}\right) & \Rightarrow &  
\left[E^{a}_{R}(n,i),E^{b}_{R}(n,i)\right] ~= ~i\epsilon^{abc} E^{c}_{R}(n,i)  
\label{ccr} 
\eea 
In (\ref{ccr}), $\sigma^{a}$ (a=1,2,3) are the Pauli matrices and the operators 
on different links commute. The angular momentum algebras on the r.h.s. of (\ref{ccr})
follows from the Jacobi identities: 
$\left[E^a_{L(R)},\left[E^b_{L(R)},U_{\alpha\beta}\right]\right]$ + cyclic permutations 
$\equiv 0$. The link operators commute amongst themselves: 
$\left[U_{\alpha\beta}(n,i),U_{\gamma\delta}(m,j)\right] = 
\left[U_{\alpha\beta}(n,i),U^{\dagger}_{\gamma\delta}(m,j)\right]=0$.  
Further,  ${E}^{a}_{L}(n,i)$ and $E^{a}_{R}(n+i,i)$, being  
the body, space fixed components of the angular momentum operator of the rigid rotator 
on the link (n,i),  mutually commute $\left[E^{a}_{L}(n,i),E^{b}_{R}(n+i,i)\right] =0$. For the 
same reason, they satisfy the kinematical constraint: 
\bea 
\sum_{a=1}^{3} {E}^{a}_{L}(n,i){E}^{a}_{L}(n,i) = 
\sum_{a=1}^{3} E^{a}_{R}(n+i,i)E^{a}_{R}(n+i,i) \equiv E^{2}(n,i) , ~~ \forall (n,i) 
\label{consn} 
\eea  
ensuring that their magnitudes are equal. The SU(2) gauge transformations correspond to 
separately rotating the body, space fixed frames of the rigid rotator \cite{kogut}: 
\bea
E_{L(R)}(n,i) \rightarrow \Lambda(n) E_{L(R)}(n,i) \Lambda^{\dagger}(n), ~~~~~~ 
U(n, i) \rightarrow \Lambda(n)U(n, i)\Lambda^{\dagger}(n+i). 
\label{gt1n}
\eea
In (\ref{gt1n}), $E_{L(R)} \equiv \sum_{a=1}^{3} E^a_{L(R)}\sigma^a$ and  $\Lambda(n)$ is 
rotation matrix in the fundamental representation of SU(2) at lattice site n. 
The commutation relations (\ref{ccr}) along with the gauge transformations (\ref{gt1n}) 
imply that the generators of SU(2) gauge transformations at any lattice site n are: 
\bea 
{\cal C}^{a}(n) = \sum_{i=1}^{3}\big(E_{L}^{a}(n,i) + E_{R}^{a}(n,i)\big), ~~\forall n,a. 
\label{su2gln} 
\eea
The corresponding Gauss law constraints are ${\cal C}^{a}(n) =0$.  
The commutation relations (\ref{ccr}) and the constraints (\ref{consn}) 
imply that  a complete set of commuting observables on every link $(n,i)$ are:
$E^{2}(n,i), {E}^{z}_{L}(n,i),$ $ E^{z}_{R}(n+i,i)$ 
where $ {E}^{z}_{L(R)}(n,i) \equiv {E}^{a=3}_{L(R)}(n,i)$ are the 
third components of the angular momenta. 
The corresponding orthonormal basis is denoted by 
~$|j(n,i),m(n,i),\tilde{m}(n+i,i)\rangle$ ~ where $j(n,i),~m(n,i)$, and $\tilde{m}(n+i,i)$ 
are the eigenvalues of the above 3 mutually commuting operators respectively.    
We now define the prepotential operators through the Jordan-Schwinger representation 
of the angular momentum algebra \cite{schwinger}: 
\bea
{E}^{a}_{L}(n,i) \equiv \frac{1}{2} a^{\dagger}(n,i){\sigma^{a}} a(n,i), ~~~~~~
E^{a}_{R}(n,i) \equiv \frac{1}{2} b^{\dagger}(n,i){\sigma}^{a} b(n,i). 
\label{sbn}
\eea
The mapping (\ref{sbn}) corresponds to associating two doublets of harmonic 
oscillator  prepotentials $a^{\dagger}_{\alpha}(n,i)$ and $b^{\dagger}_{\alpha}(n+i,i)$ 
and their conjugates to the initial and the end points of the link $(n,i)$ respectively.     
This assignment is shown in  Figure (\ref{fig2}b).  
The canonical electric field or angular momentum commutation relations in (\ref{ccr}) are 
satisfied provided the prepotentials on every link $(n,i)$ satisfy the standard harmonic 
oscillator algebra: 
\bea 
\left[a_{\alpha},a^{\dagger}_{\beta}\right] = \delta_{\alpha\beta}, 
~~~~\left[b_{\alpha},b^{\dagger}_{\beta}\right] = \delta_{\alpha\beta}, 
~~~~ \left[a_{\alpha},a_{\beta}\right]=0, 
~~~~ \left[b_{\alpha},b_{\beta}\right]=0.  
\label{ftg} 
\eea
The body and space fixed components of the angular momentum or electric fields mutually commute 
implying: $[a_{\alpha},b^{\dagger}_{\beta}] =  \left[a_{\alpha},b_{\beta}\right] =0$. 
Note that the prepotential vacuum state $|0\rangle$ on the link (n,i) satisfying: 
$a_{\alpha}(n,i)|0\rangle =0,~ b_{\alpha}(n,i)|0\rangle =0$ is the strong coupling vacuum  
defined as: $E_L^a(n,i)|0 \rangle =0,~ E_R^a(n,i)|0 \rangle =0$. In \cite{wiese} anti-commuting 
oscillators, instead of (\ref{ftg}),  are used in (\ref{sbn}) to treat QCD as 
quantum link models.  We  define the total number operators: 
\bea 
N_a(n,i) = a^{\dagger}(n,i) \cdot a(n,i),~~~~~~  N_b(n,i) = b^{\dagger}(n,i) \cdot b(n,i), 
\nonumber 
\eea
where $ a^{\dagger} \cdot a \equiv \sum_{\alpha=1}^{2} a_{\alpha}^{\dagger}a_{\alpha}$. 
The left and right Casimirs are: $E_{L}\cdot E_{L} \equiv \sum_{a=1}^{3} E^a_{L}E^a_{L}=
N_{a}/{2}\left(N_a/2+1\right),$ $   
E_{R}\cdot E_{R} =N_{b}/2\left(N_b/2+1\right)$. 
Therefore, the kinematical constraints (\ref{consn}) in terms of the prepotentials mean 
that on every link $(n,i)$, the number of left oscillators is equal to the number of right 
oscillator: 
\bea 
N_a(n,i)= N_b(n+i,i) \equiv N(n,i)
\label{noe}
\eea 
as shown in  Figure (\ref{fig2}b). 
Under SU(2) gauge transformations (\ref{gt1n}) at site n, the defining equations (\ref{sbn}) 
imply that the prepotentials transform as SU(2) fundamental doublets: 
\bea
a_{\alpha}(n,i) \rightarrow \Lambda_{\alpha\beta}(n) a_{\beta}(n,i), ~~~~~~
b_{\alpha}(n,i) \rightarrow \Lambda_{\alpha\beta}(n) b_{\beta}(n,i). 
\label{gt3n}
\eea
Therefore, unlike the link operators $U_{\alpha\beta}(n,i)$ transforming at both the ends 
of the link by $\Lambda(n)$ and $\Lambda(n+i)$ in (\ref{gt1n}), the prepotentials 
$\left(a_{\alpha}(n,i), b_{\alpha}(n,i)\right)$ transform only at a single end by 
$\Lambda(n)$. In the next section this simple fact will enable us to define 
SU(2) invariant Hilbert spaces ${\cal H}^{SU(2)}(n)$ locally at every lattice site n. 

In terms of the prepotential operators we have additional U(1) gauge invariance 
as their defining equations (\ref{sbn}) are invariant under: 
\bea 
a^{\dagger}_{\alpha}(n,i) \rightarrow \lambda(n,i) ~ a^{\dagger}_{\alpha}(n,i); ~~  
b^{\dagger}_{\alpha}(n+i,i) \rightarrow \lambda^{\dagger}(n,i) ~b^{\dagger}_{\alpha}(n+i,i).   
\label{u1nn}
\eea 
In (\ref{u1nn}), $\lambda(n,i) = exp i \theta(n,i)$ is the abelian phase factor on the 
link $(n,i)$. Note that the abelian transformations (\ref{u1nn}) leave the Hilbert space 
of the rigid rotators, satisfying the constraints (\ref{noe}), unchanged. 
The  creation operators $a^{\dagger}_{\alpha}(n,i)$ and $b^{\dagger}_{\beta}(n+i,i)$ 
create and absorb unit abelian flux at lattice sites n and n+i on the link (n,i) respectively. 
Thus the operators $a^{\dagger}_{\alpha}(n,i)b^{\dagger}_{\beta}(n+i,i), ~ 
a_{\alpha}(n,i)b_{\beta}(n+i,i)$ create and annihilate an abelian flux line on the link 
$(n,i)$ respectively. To analyze nonabelian SU(2) fluxes in terms of the prepotential 
operators, we consider the following commutation relations:
\bea 
\left[E_{L}^a,a^{\dagger}_{\alpha}\right] = 
\frac{1}{2} a^{\dagger}_{\beta} \sigma^{a}_{\beta\alpha},~~~~
\left[E_{R}^a,b^{\dagger}_{\alpha}\right] = 
\frac{1}{2} b^{\dagger}_{\beta} \sigma^{a}_{\beta\alpha}. \nonumber 
\eea
The above relations imply that $a^{\dagger}_{\alpha}(n,i) b^{\dagger}_{\beta}(n,i)$ 
and $a_{\alpha}(n,i) b_{\beta}(n,i)$ on a link (n,i), like the the link operators 
$U_{\alpha\beta}(n,i)$, change the angular momentum $j(n,i)$ by $ \pm\frac{1}{2}$. 
As an example the 
prepotential operators acting on 
the vacuum with $j = 0$ create $j=\frac{1}{2}$ state: 
$\left(E_L \cdot E_L\right) \left[a^{\dagger}_{\alpha} | 0 \rangle \right]= 
\frac{3}{4}\left[a^{\dagger}_{\alpha} | 0 \rangle\right], ~~
\left(E_R \cdot E_R\right) \left[b^{\dagger}_{\alpha} | 0 \rangle \right] = 
\frac{3}{4}\left[b^{\dagger}_{\alpha} | 0 \rangle \right]$. These results correspond to 
$\left(E_L \cdot E_L\right) \left[U_{\alpha\beta} | 0 \rangle \right] = 
\frac{3}{4}\left[ U_{\alpha\beta} | 0 \rangle\right]$ and $\left(E_R \cdot E_R\right) 
\left[U_{\alpha\beta} | 0 \rangle \right]  = \frac{3}{4}\left[U_{\alpha\beta} | 0 \rangle \right]$.  
More precisely, using the Wigner-Eckart theorem, the link operator can also be represented 
in terms of the prepotential operators \cite{manu1}: 
\bea
U_{\alpha\beta}(n,i)  =  F(n,i) \big(\tilde{a}^{\dagger}_{\alpha}(n,i) {b}^{\dagger}_{\beta}(n+i,i)
+{a}_{\alpha}(n,i) \tilde{b}_{\beta}(n+i,i)\big) F(n,i).
\label{dhn}
\eea
In (\ref{dhn}), $F(n,i) \equiv \left(N(n,i)+1\right)^{-\frac{1}{2}}$   
is the normalization factor where $N(n,i)$  is defined in (\ref{noe}). 
Note that the r.h.s. of (\ref{dhn}) is U(1) invariant and also has 
the required SU(2) gauge transformation property of $U(n,i)$ given in 
(\ref{gt1n}). Acting on the Hilbert space satisfying (\ref{noe}), 
the relation (\ref{dhn}) is consistent with (\ref{det1}), (\ref{ccr}) and 
$[U_{\alpha\beta},U_{\gamma\delta}]=[U_{\alpha\beta},U^{\dagger}_{\gamma\delta}] 
=0$. The relations (\ref{sbn}) and (\ref{dhn}) 
provide complete mapping from the original operators satisfying (\ref{ccr})  
and the constraints (\ref{consn}) to the harmonic oscillator prepotential operators satisfying 
(\ref{ftg}) and the constraints (\ref{noe}) respectively.  For later convenience, we define the 
following  $\pi$ operation on every link: 
\bea 
\tilde{a}^{\dagger}_{\alpha}(n,i)  
~\overset{\pi}{\rightarrow}~ 
{a}_{\alpha}(n,i),~~~~~
{b}^{\dagger}_{\alpha}(n,i) 
~{\overset{\pi}{\rightarrow}} ~
\tilde{b}_{\alpha}(n,i) 
\label{pari} 
\eea
The link operator can now  be written as $U_{\alpha\beta}(n,i)  =  
F(n,i) \left(\tilde{a}^{\dagger}_{\alpha}(n,i) {b}^{\dagger}_{\beta}(n+i,i) 
+ \pi(n,i)\right)F(n,i)$.  
\begin{figure}[t]
\begin{center}
\includegraphics[width=0.35\textwidth,height=0.3\textwidth]
{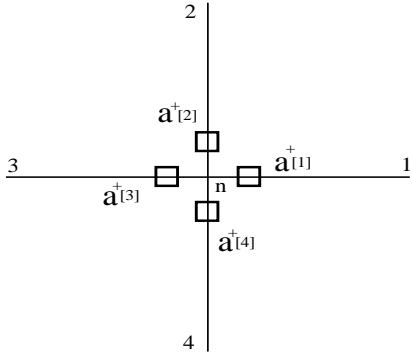}
\end{center}
\vspace{-5mm}
\caption{The 2d prepotential SU(2) doublets $a^{\dagger}[n,i]$, i=1,2,...,2d
around every lattice site n shown in d = 2 by their Young tableau boxes 
$\square$.  They all transform as doublets under SU(2) gauge transformation 
at site n.}  
\label{fig3}
\end{figure} 

\subsection{Non-abelian intertwining, abelian weaving and loop states} 

\begin{figure}[t]
\begin{center}
\includegraphics[width=0.85\textwidth,height=0.2\textwidth]
{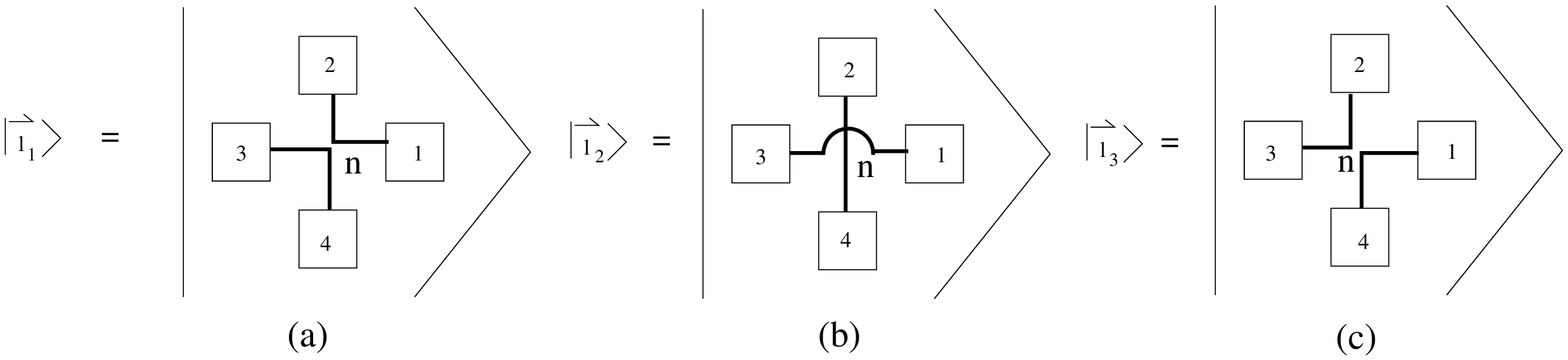} 
\end{center}
\vspace{-5mm}
\caption{Graphical representation of SU(2) invariant intertwining illustrated for 
the states $|\vec{l}_1 ~\rangle,  |\vec{l}_2 ~\rangle$ and $|\vec{l}_3 ~\rangle$. 
The constraint  $|\vec{l}_1 ~\rangle = |\vec{l}_2 ~\rangle  - 
|\vec{l}_3 ~\rangle$ at lattice site n represent the Mandelstam constraint in Figure (\ref{fig2}) 
in terms of the prepotentials. The thick lines 
should be compared with the corresponding thick lines in Figure 
(\ref{fig1}).} 
\label{fig4}
\end{figure}

The advantage of the prepotential operators is that under SU(2) gauge transformations 
they transform locally as SU(2) fundamental doublets (\ref{gt3n}). Therefore, the  SU(2) 
invariant loop Hilbert space ${\cal H}^{L}$ can be constructed and analyzed {\it locally} 
in terms of ${\cal H}^{SU(2)}(n)$ at different lattice sites n. In the next 
section this simple fact, in turn, will enable us to solve the Mandelstam constraints exactly. 
To appreciate and elaborate on these statements further, it is convenient to relabel all 
the prepotentials and the corresponding electric fields around a lattice site n. 
We define\footnote{Unless stated explicitly, the 
direction indices $[i],[j].. $ within square brackets will vary from 1 to 2d.}: 
$a^{\dagger}_{\alpha}[n,{i}],~ {i} =1,2,...,2d$ 
where $a^{\dagger}_{\alpha}[n,i] \equiv a^{\dagger}_{\alpha}(n,i),~ a^{\dagger}_{\alpha}[n,d+i] 
\equiv b^{\dagger}_{\alpha}(n,i), i=1,2,..,d$ as shown in Figure (\ref{fig3}). 
Thus, instead of the original $\left[E_{L}^{a}(n,i),E_{R}^{a}(n,i),U_{\alpha\beta}(n,i)\right]$ 
description, we now have $2d$ SU(2) doublet prepotential operators 
around every lattice site n as shown in Figure (\ref{fig3}). 
The SU(2) gauge transformation:  
\bea 
a_{\alpha}^{\dagger}[n,i] \rightarrow a_{\beta}^{\dagger}[n,i] \Lambda^{\dagger}_{\beta\alpha}(n), ~~
i=1,2,..,2d. 
\nonumber 
\eea 
is represented in Figure (\ref{fig3}) by a single Young tableau box on the link $[n,i]$ with 
$i=1,2,..2d$. Further  defining  $J^a[n,i] = a^{\dagger}[n,i] \frac{\sigma^a}{2} a[n,i]$, we get:
\bea 
J^{a}[n,i] = E^{a}_{L}(n,i); ~~~J^{a}[n,d+i] = E^{a}_{R}(n,i); ~~~~i=1,2,...,d.
\label{nsa}    
\eea
Therefore, all possible SU(2) invariant operators at site n are given by 
``intertwining" (anti-symmetrizing) any two different prepotential SU(2) 
doublets:  
\bea 
L_{ij}(n) = \epsilon_{\alpha\beta}~  a^{\dagger}_{\alpha}[n,i]~ {a}^{\dagger}_\beta[n,j]
\equiv a^{\dagger}[n,i] \cdot\tilde{a}^{\dagger}[n,j], ~~~~~ i,j=1,2,...,2d; ~~i < j. 
\label{int} 
\eea
In (\ref{int}), $\epsilon_{\alpha\beta}$ is completely antisymmetric tensor
($\epsilon_{11}=\epsilon_{22}= 0, \epsilon_{12}=-\epsilon_{21}= 1$) and
$\tilde{a}^{\dagger}_{\alpha} \equiv \epsilon_{\alpha\beta}
{a}^{\dagger}_{\beta}$.  The $d(2d-1)$ intertwining operators $L_{ij}(n)$ in (\ref{int}) 
correspond to  putting  the Young boxes in $[i]$ and $[j]$  directions vertically to construct SU(2) 
singlets. In Figure (\ref{fig4}), we graphically show this by joining these boxes with thick lines.   
These intertwining operators along with $a^{\dagger}[n,i]\cdot a[n,j]$ are the basic SU(2) invariant 
operators. Any  gauge invariant operator can be analyzed {\it locally} in terms of 
them\footnote{See section 3 and 
appendix B for an explicit analysis of $Tr U_{\rm plaquette}$.}. Further, the $d(2d-1)$ 
operators $L_{ij}(n)$ at lattice site n play the role of ``SU(2) gauge invariant bricks"  which 
acting on the strong coupling vacuum create the complete SU(2) gauge invariant Hilbert space 
${\cal H}^{SU(2)}(n)$ at n. Note that $L_{ij}(n) = -L_{ji}(n)$, $L_{ii}=0$ implying  self intertwining is
not allowed. Thus  a basis in ${\cal H}^{\rm SU(2)}(n)$  
is given by: 
\bea
|\vec{l}(n)~ \rangle  \equiv \left\vert \begin{array}{cccccc}
l_{12} & l_{13}  & l_{14}& ....& l_{1(2d)}     \\
 & l_{23} & l_{24} & .... &l_{2(2d)} \\
 &  &  .... & .... &.... \\
 &  &  & l_{2d-2(2d-1)} & l_{2d-2(2d)}  \\
 &  &  &  & l_{2d-1(2d)}   \\
\end{array} \right \rangle =  \prod_{{}^{{i},{j}=1}_{{j}  \rangle 
{i}}}^{2d} \left(L_{ij}(n)\right)^{l_{ij}(n)}
|0 \rangle ,  ~~ l_{ij}(n) \in {\cal Z_{+}}.
\label{giv2n}
\eea
In (\ref{giv2n}), ${\cal Z_{+}}$ denotes the set of all non-negative integers and 
$l_{ij} (\equiv l_{ji}, l_{ii} = 0)$  are $d(2d-1)$ SU(2) gauge invariant intertwining 
integer quantum numbers characterizing the SU(2) gauge invariant Hilbert space at the 
site n. 
These SU(2) singlet states can also be graphically represented by first 
assigning $l_{ij}(n)$ Young tableau boxes individually to the links $[n,i]$ 
and $[n,j]$  and then joining (intertwining) them together.  
In Figure (\ref{fig4}) we illustrate this graphical representation 
for the simple case involving the following three vectors in $d=2$: 
\bea 
|\vec{l}_1 ~\rangle   = \left\vert \begin{array}{cccc} 
1 & 0  & 0     \\
 & 0 & 0\\
 &  &  1\\
\end{array} \right \rangle, ~~~~ 
|\vec{l}_2 ~\rangle  = \left\vert \begin{array}{cccc} 
0 & 1  & 0     \\
 & 0 & 1\\
 &  & 0\\
\end{array} \right \rangle, ~~~~ 
|\vec{l}_3 ~\rangle  = \left\vert \begin{array}{cccc} 
0 & 0  & 1     \\
 & 1 & 0\\
 &  &  0\\
\end{array} \right \rangle   
\label{3v}
\eea
The above three states $|\vec{l}_1 ~\rangle = L_{12}L_{34}|0\rangle, ~|\vec{l}_2 ~\rangle = 
L_{13}L_{24}|0\rangle, ~|\vec{l}_3 ~\rangle= L_{14}L_{23}|0\rangle$   
are manifestly invariant under SU(2) gauge transformation. 
More explicitly, $|\vec{l}(n) ~\rangle $ in (\ref{giv2n}) satisfy the SU(2) 
Gauss law (\ref{su2gln}) constraints:
\bea 
{\cal C}^{a}(n) |{\vec{l}}(n) ~\rangle  = \sum_{i=1}^{2d} J^{a}[n,i] |{\vec{l}}(n)~ \rangle 
=J^{a}_{total} |{\vec{l}}(n)~ \rangle = 0 
\label{glnn}  
\eea
as $[{\cal C}^{a}(n),L_{ij}(n)] = 0$ and ${\cal C}^{a}(n) |0\rangle =0$.  Thus 
all the states $|\vec{l}(n)~\rangle$ in (\ref{giv2n}) at site n are the eigenstates 
of the total angular momentum at n with eigenvalues zero. Infact, the states $|\vec{l}(n) ~\rangle$ 
are also eigenstates of individual $2d$ Casimirs in (\ref{nsa}): 
\bea 
J[n,i].J[n,i] |{\vec{l}}(n)~ \rangle  
 = j[n,i](j[n,i]+1) |{\vec{l}}(n) ~\rangle , ~~ i=1,2,3,4 
\label{eve2} 
\eea
where, 
\bea 
2j[n,i] = \sum_{k\neq i=1}^{2d} l_{ik}(n),~~ l_{ik}(n)=l_{ki}(n),~~ l_{ik}(n) \in {\cal Z_{+}}  
\label{part} 
\eea
We note that (\ref{part}) is both necessary and sufficient condition on $j[n,i], 
i=1,2,3,..,2d$ to get SU(2) singlets.  

Having solved the SU(2) Gauss law, we now 
focus on the abelian gauge transformations (\ref{u1nn}). As shown in Figure (\ref{fig2}b), 
the  abelian Gauss law (\ref{noe}) states that on any link the number of 
{\bf a} type oscillators at the left end is equal to the number of {\bf b} type 
oscillators  at the right end. This can again be easily satisfied 
by putting  $N_{a}(n,i)=N_{b}(n,i) = N(n,i)$ abelian flux lines on 
every link (n,i). Therefore, geometrically, the $SU(2)\otimes U(1)$ Gauss law   
demands the continuity of these U(1) flux lines on the entire lattice through 
SU(2) intertwining at every lattice site.  In other words, all possible SU(2) 
invariant intertwining within ${\cal H}^{\rm SU(2)}(n)$ and U(1) 
weaving of the neighboring ${\cal H}^{\rm SU(2)}(n)$ is geometrically equivalent 
to  drawing all possible loops on the lattice leading to the loop Hilbert space 
${\cal H}^L$. Given a configuration of closed 
loops on a lattice one can read off all the intertwining quantum numbers 
$l_{ij}(n)$ at site n by simply counting the number of loop lines going from 
$[i]^{th}$ to $[j]^{th}$ direction. The number of loops passing through a 
link $(n,i)  = N_a(n,i) =2j(n,i) = N_b(n+i,i)$ where $2j(n,i)$ is given in 
(\ref{part}).   
The reverse is also true: given $l_{ij}(n) ~ \forall~ n$, which are consistent with the 
U(1) gauge invariance, one can always draw corresponding closed loops. 
We now review the Mandelstam constraints in the basis (\ref{giv2n}) before solving 
them explicitly in terms of the prepotential operators.  

\subsubsection{Mandelstam constraints revisited}

The basis $|\vec{l}(n)~\rangle$ in (\ref{giv2n}) provides the  
local description  of the Wilson loop basis in 
terms of the intertwining operators and intertwining quantum numbers at lattice site n.   
To see this explicitly, one writes  Wilson loop operator ${\rm Tr}~ U_{\gamma}$ 
corresponding to a loop $\gamma$ in terms of the prepotential operators using 
(\ref{dhn}).  It is clear that acting on the strong coupling vacuum they will 
produce states of the form $|\vec{l}(n) ~\rangle$ at every lattice site traversed 
by the loop $\gamma$. Thus, following the notation of section (2.1),  the the Wilson 
loop basis can be locally characterized 
by ${\cal M}/n^{d}$ =  (Number of intertwining quantum numbers per site $ - $ Number of U(1) 
constraints per site) $= d(2d-1) -d = 2d(d-1)$ integers per lattice site. 
This establishes that the basis $|\vec{l}(n)~\rangle$ describing the Wilson 
loop basis locally in terms of intertwining quantum numbers is over-complete. To 
illustrate this further and more explicitly, we again consider the vectors $|\vec{l}_1~\rangle,~ 
|\vec{l}_2~\rangle,~|\vec{l}_3~\rangle$ in (\ref{3v}) and shown in Figure (\ref{fig4}).
Using the identity:
\bea   
(a^{\dagger} \cdot \tilde{b}^{\dagger})(c^{\dagger} \cdot \tilde{d}^{\dagger})  
\equiv  (a^{\dagger} \cdot \tilde{c}^{\dagger})(b^{\dagger} \cdot \tilde{d}^{\dagger}) 
- (a^{\dagger} \cdot \tilde{d}^{\dagger})(b^{\dagger} \cdot \tilde{c}^{\dagger}),   
\label{mandi}
\eea 
we find the vectors in (\ref{3v}) are linearly dependent: 
\bea 
|\vec{l}_1~ \rangle  = |\vec{l}_2 ~\rangle ~ -~ |\vec{l}_3 ~\rangle  
\label{nli}
\eea  
\noindent Infact, the SU(2) identity (\ref{mc}) involving link operators 
corresponds to the identity (\ref{mandi}) and the Mandelstam constraint (\ref{3sr}) is 
the constraint (\ref{nli}) written in terms of the  relevant 
prepotential operators at site n. This can again be seen 
by writing the Mandelstam identity (\ref{mc}) 
in terms of prepotentials using (\ref{dhn}) or  
comparing Figure (\ref{fig1}a), (\ref{fig1}b) and (\ref{fig1}c)  with 
Figure (\ref{fig4}a), (\ref{fig4}b) and (\ref{fig4}c) respectively.  
In the prepotential language, the identities  of the form 
(\ref{mandi}) and  their various powers 
will make the $|\vec{l} ~\rangle $ basis in (\ref{giv2n}) linearly dependent. 
Infact, this is what was done in (\ref{mcc}) to get all the loop states on the 
two plaquettes related by Mandelstam constraints. Thus, at this stage, 
the problem of over-completeness of the Wilson loop basis can be analyzed 
locally in terms of the prepotentials.    

\subsubsection{The solutions} 

To solve the Mandelstam constraints we now need to focus only on a single lattice 
site\footnote{From now onwards we will be  working locally at a given lattice 
site. Therefore, we will ignore the site index unless necessary. Also, we use the 
notation: $J_{i}^2 \equiv \sum_{a=1}^{3}J^{a}[n,i]J^{a}[n,i], i=1,2...,2d$, the 
eigenvalues of $J_{i}^2$ are denoted by $j_i(j_i+1)$.} n. As discussed in section 
(2.2), the initial complete set of commuting observables at n 
consists of $4d$ angular momentum operators: $J_i^2,J_i^{z}, i=1,2,..,2d$. Instead, we can also 
consider an equivalent basis where the following 4d mutually commuting operators 
\cite{sharat1,sharat2} are diagonal: 
\bea 
CSCO \equiv \left[J_1^{2},...J_{2d}^2; \left(J_1+ J_2\right)^{2},
\left(J_1+J_2+J_3\right)^2 ..., \left(J_1+ J_2+J_3...... J_{(2d-1)}\right)^{2},
 J_{total}^{2}, J_{total}^{z}\right] 
\label{amm} 
\eea   
where $J_{total}^{2} = \left(J_1+ J_2^{2}+J_3......+J_{2d}\right)^{2}$ 
and $J_{total}^{z} = \left(J^{a=3}_1+ J^{a=3}_2+J^{a=3}_3......+J^{a=3}_{2d}\right)$. 
The SU(2) Gauss law (\ref{glnn}) demands  $J_{total}^{2}= J_{total}^{z} =0$.  
Therefore, we drop the last two total angular momentum operators from the list (\ref{amm}). 
For later analysis, it is convenient to divide the remaining $(4d-2)$ operators in 
$CSCO$ (\ref{amm}) into two parts: 
\bea 
CSCO(I) = \left[J_1^{2}, J_2^{2},..,J_{2d}^2\right],~~~ CSCO(II) =
\left[\left(J_1+J_2\right)^2, ...,\left(J_1+ J_2+J_3...... J_{(2d-1)}
\right)^{2} = J_{2d}^2\right] 
\label{amm2} 
\eea     
The CSCO(I) contains 2d angular momentum Casimir operators along the 2d directions and  
the CSCO(II) contains the remaining $(2d-2)$ Casimirs in the above chosen angular momentum 
addition scheme.  On ${\cal H}^{SU(2)}(n)$ the last two operators in CSCO(II) are equal 
because of the SU(2) Gauss law (\ref{glnn}). We can, therefore, denote the corresponding 
SU(2) gauge invariant orthonormal eigenvectors by \cite{sharat1,sharat2}  
\bea 
|j_1,j_2,..j_{2d};j_{12},j_{123},....,j_{12..(2d-1)}=j_{2d}\rangle 
\equiv |j_1,j_2,j_{12},j_3,j_{123},.....,j_{2d-1},j_{12..(2d-1)}=j_{2d}\rangle.    
\label{amcl} 
\eea
The states in (\ref{amcl}) are characterized by  maximum possible $(4d-3)$ 
``good quantum numbers" which can be 
simultaneously measured at every lattice site. The SU(2) gauge invariant states 
$|\vec{l}~\rangle$ in (\ref{giv2n}) are already eigenstates of of CSCO(I) 
with eigenvalues $2j_i =\sum_{k=1}^{2d}l_{ik}$ (see (\ref{eve2})). 
Therefore,  we can relabel them in terms of their angular momenta: 
\bea 
|~\vec{l}~\rangle \rightarrow |j_1,j_2,....,j_{2d},j_{total}=m_{total} 
=0\rangle
\label{lj}
\eea
We note that the mapping (\ref{lj}) is many to one or degenerate because of the following 
discrete symmetries of the angular momenta $j_i$ (i=1,2,...,2d) in (\ref{part}): 
\bea 
&&l_{i_1i_2} \rightarrow l_{i_1i_2} + r + t,~~ l_{i_1i_3} \rightarrow l_{i_1i_3} -r +s,~~ 
l_{i_1i_4} \rightarrow l_{i_1i_4} -s -t, \nonumber \\ 
&& l_{i_2i_3} \rightarrow l_{i_2i_3} -s -t,~~ l_{i_2i_4} \rightarrow l_{i_2i_4} -r +s,~~ 
l_{i_3i_4} \rightarrow l_{i_3i_4} +r +t.
\label{symm}
\eea
In (\ref{symm}) $i_1 \neq i_2 \neq i_3 \neq i_4$ are any four different directions and 
r,s and t can take all possible $\pm$ integer values such that $l_{i,j} \ge 0$.
The advantage of the mapping (\ref{lj}) is that the  states in (\ref{giv2n}) with different 
$\vec{l}$ which are related by (\ref{symm}) are precisely the states related by the the 
Mandelstam 
constraints (\ref{mandi}). In other words, the states related by Mandelstam constraints are 
degenerate with respect to CSCO(I).  The reverse is also true: the degenerate states with 
respect to CSCO(I) are all related by the Mandelstam constraints (\ref{mandi}).  As an 
example, we again consider the three loop states in (\ref{3v}): $|\vec{l}_1 ~\rangle , 
|\vec{l}_2 ~\rangle $ and $|\vec{l}_3~ \rangle $ in d=2  which are related by the Mandelstam 
constraints (\ref{mandi}). In terms of the angular momenta: 
\bea 
\left. \begin{array}{cc} 
|\vec{l}_1 ~\rangle\\
|\vec{l}_2 ~\rangle\\
|\vec{l}_3 ~\rangle\\ \end{array}\right\}  \rightarrow  |j_{i_1}=j_{i_2}=j_{i_3}=j_{i_4} =\frac{1}{2}, 
j_{total}=m_{total}=0 \rangle. 
\nonumber 
\eea
Also, given $j_{1}=j_{2}=j_{3}=j_{4} =\frac{1}{2}$, the above three possible partitions 
$\{\vec{l}_1\}, \{\vec{l}_2\}$ and $\{\vec{l}_3\}$ given in (\ref{3v}), are mutually related 
by (\ref{symm}).  Therefore, we  can  lift the degeneracy and solve the Mandelstam 
constraints by demanding that the CSCO(I) degenerate eigenbasis (\ref{giv2n}) 
to be the eigenstates of CSCO(II) as well. We conclude that a complete 
orthonormal loop basis in d dimension is locally characterized by
$(4d-3)$ angular momentum quantum numbers and are given in (\ref{amcl}). 
Note that the $(2d-3)$ eigenvalues of CSCO(II) are not free 
and have to satisfy the triangular constraints: 
\bea 
|j_{12..(k-1)}-j_k| \le j_{12..k} \le j_{12..(k-1)}+j_k, ~~k=2,3,...,(2d-1)  
\label{cont} 
\eea
along with $j_{12..(2d-1)}=j_{2d}$. 

\subsubsection{Solving the triangular constraints} 

The above characterization of the physical Hilbert space  
in terms of the angular momentum quantum numbers has been 
given in the context of duality transformation in 
lattice gauge theories \cite{sharat1} leading to a description 
in terms of triangulated surfaces \cite{sharat2}. 
In this section, we further solve the triangular constraints,  geometrically 
representing triangulated 2 dimensional surfaces, in terms of the 
intertwining quantum numbers $\{\vec{l}\}$ which geometrically represent one dimensional 
loops. 

In terms of the prepotentials (see Figure (\ref{fig3})) the 
SU(2) invariant  states $|j_1,j_2,...,j_{2d}\rangle$ in the mapping 
(\ref{lj}) represent  $ \sum_{k}l_{ik} (= 2j_i)$ Young tableau 
boxes on the $[i]^{th}$ link. Let us first consider (12) plane. 
To get the state $|j_1,j_2,...,j_{2d};j_{12}\rangle$ from the degenerate state $|j_1,j_2,...,j_{2d}\rangle$, 
we need to intertwine (antisymmetrize) $l_{12}$ boxes from $2j_1$ boxes with $l_{12}$ 
boxes from $2j_2$ boxes so that we are left with $2j_{12}$ boxes in the (12) plane. 
Therefore, $2j_{12} = 2j_1+2j_2-2l_{12}$. This process is sequential  
and can be repeated to get the eigenvalues of the 
CSCO(II) also in terms of the linking numbers: 
\bea 
l_{12} & = & j_{1}+j_{2}-j_{12} \nonumber \\
l_{13}+l_{23}  & = & j_{12}+j_{3}-j_{123} \nonumber \\
l_{14}+l_{24}+l_{34} & = & j_{123}+j_{4}-j_{1234} \nonumber \\
........ & &........ \nonumber \\
l_{1(2d)}+l_{2(2d)}+...+l_{1(2d)} & = & j_{12...(2d-1)}(=j_{2d})+j_{2d}
- j_{12...(2d)}(=0)  = 2j_{2d} 
\label{cln} 
\eea 
Given $j_{12}$ at a lattice site, the top equation fixes $l_{12}$, the next line 
fixes $l_{13}+l_{23}$ in terms of $j_{12}$ and $j_{123}$,........, the last equation 
in (\ref{cln}) is an identity as it is already contained in (\ref{part}). 
In appendix A, a detailed technical calculation, involving properties of SU(2) 
coherent states, shows that the final orthonormal and manifestly 
SU(2) gauge invariant loop states $|LS\rangle_{n}$ spanning the  Hilbert space 
${\cal H}^{SU(2)}$(n) are:
\begin{eqnarray}
\label{std2}
|LS\rangle_{n} \equiv  |j_1,j_2,..j_{2d};j_{12},j_{123},...j_{12..(2d-1)}=j_{2d} \rangle
 =  N(j) \sum_{\{l\}}\hspace{-0.05cm}
{}^{{}^\prime} \prod_{{}^{i,j}_{i < j}}
\frac{1}{l_{ij}!}
\big({L}_{ij}(n)\big)^{l_{ij}(n)} |0 \rangle 
\end{eqnarray}
The prime over the summation means that,  the linking numbers $l_{ij}$ are
are summed over all possible values which are consistent with (\ref{part}) 
and (\ref{cln}).  {\it This summations corresponds to taking appropriate linear 
combination of the degenerate loop eigenstates of CSCO(I) related by the Mandelstam 
constraints to produce an orthonormal and complete basis.} 
The normalization constant in (\ref{std2}) is given by:  
\bea 
 N(j)  =  {N(j_1,j_2,j_{12})}{N(j_{12},j_{3},j_{123}})
N(j_{123},j_4,j_{1234})......{N(j_{2d},j_{2d},0)} 
\label{nnaa} 
\eea
where $N(a,b,c)  =  \Big[\frac{(2c+1)}{(a+b+c+1)!}\Big]^{\frac{1}{2}}
\Big[{(-a+b+c)!(a-b+c)!(a+b-c)!}\Big]^{\frac{1}{2}}.$
We emphasize that this simple 
construction (\ref{std2}) of the orthonormal loop basis in terms of intertwining 
operators $L_{ij}$ and intertwining linking numbers $l_{ij}$ in arbitrary dimensions 
becomes  extremely involved and complicated in terms of the link operators $U_{\alpha\beta}$ and the 
angular momentum quantum numbers\footnote{The 
states $|j_1,j_2,j_{12},j_3,j_{123}, ....,j_{2d-1},j_{12..(2d-1)},j_{total},m_{total}\rangle$ 
can be obtained  by using Clebsch-Gordan coefficients: 
$|j_1,j_2,j_{12},j_3,j_{123},...j_{12..(2d-1)}=j_{2d} \rangle$ 
= $\sum_{\vec{m}}$   
$C_{j_{12..(2d-1)}m_{12..(2d-1)},j_{2d}m_{2d}}
^{j_{12..2d}=0m_{12..2d}=0}.....$  
$C_{j_{12}m_{12},j_3m_3}^{j_{123}m_{123}} 
C_{j_1m_1,j_2m_2}^{j_{12}m_{12}}   
 \prod_{i=1}^{2d} \otimes |j_im_i\rangle. $ 
However, such approaches lead to rapid proliferation of gauge 
non-invariant Clebsch Gordan coefficients \cite{burgio} forcing one to 
use graphical methods to avoid this problem. In contrast, the construction 
(\ref{std2}) in terms of gauge invariant intertwining 
numbers (not angular momentum) is simple and bypasses this problem.}. 

As an example, we again consider the states $|\vec{l}_{1} ~\rangle , 
|\vec{l}_{2}~ \rangle $ and 
$|\vec{l}_{3}~ \rangle$ and solve the Mandelstam constraint (\ref{nli}).  
Our result (\ref{std2}) immediately gives us 
the corresponding two independent (orthonormal) states: 
\bea 
|j_1=j_2=j_3=j_4 =\frac{1}{2},j_{12}=0 \rangle   & = &  \frac{1}{2} |\vec{l}_{1} \rangle , 
\nonumber 
\\ 
|j_1=j_2=j_3=j_4 =\frac{1}{2},j_{12}=1  \rangle  &  = & \frac{1}{2\sqrt{3}} \left[|\vec{l}_{2} \rangle  + 
|\vec{l}_{3} \rangle \right]     
\eea
It can be explicitly checked that the above two states are individually normalized and 
mutually orthogonal. 

It is perhaps worth going back  and also solve 
the Mandelstam constraints for the example given in section 2.1 with arbitrary $N_{A}$ 
and $N_{B}$ in 2 dimension. In this special case: $2j_1(n)=2j_2(n)= N_A$ and $2j_3(n) 
=2j_4(n) = N_B$. The single state shown in the top r.h.s. of (\ref{mcc}) correspond 
to $l_{12}= N_{A}$ and hence $j_{12} = j_1+j_2-l_{12} =0$ at site n. The two states 
in the second top 
line of (\ref{mcc}) represent $l_{12}(n) = N_{A}-1$ and therefore correspond to $j_{12}(n)
 = 1$, the three states in the next line have $l_{12}(n) = N_A-2$ or equivalently 
$j_{12}(n)=2$  and so on so forth.  The $(N_{\rm{min}}+1)$ states in the 
last line of (\ref{mcc}) correspond to $l_{12}(n) = 0$ if $N_A<N_B$ and  $l_{12}(n) 
= N_B$ if $N_B < N_A$.  Therefore, these states at n correspond to  
$j_{12}(n)= 2j_{1}$ if $(N_A<N_B)$ and $j_{12}  = 2(j_1-j_2)$ if $(N_B<N_A)$.  
Using the relative weights in (\ref{std2}), one can easily construct an 
orthonormal basis in ${\cal H}^{\rm{SU(2)}}$(n).   

As mentioned before, the $SU(2) \otimes U(1)$ invariant orthonormal loop states 
on the lattice can be obtained by drawing all possible loops on the lattice and computing   
the $4d-3$ ``good quantum numbers" in (\ref{amcl}) and constructing (\ref{std2}) 
at every lattice site.  All possible orthonormal loop states on lattice can be 
formally written  as: 
\bea 
|LS \rangle_{\rm lattice}  = 
\left(\prod_{(n,i) \in {\rm lattice}} \int_{0}^{2\pi} 
d\phi(n,i) exp ~i ~\phi(n,i)\left(j[n,i] 
- j[n+i,d+i]\right)\right) ~ \prod_{n \in {\rm lattice}}  
\otimes |LS \rangle_{n}
\label{lsl}
\eea
The auxiliary angular fields $\phi(n,i), 0< \phi(n,i) \le 2\pi$ in (\ref{lsl}) on 
every link implement the U(1) Gauss law (\ref{noe}). Note that the  U(1) Gauss law 
provides d relations per lattice site. 
Therefore, without any over counting, there are ${\cal N}/n^d = 4d-3-d = 3(d-1)$  ``good 
quantum numbers" associated with the loop states per lattice site in the final construction (\ref{lsl}). 
This was the desired number we started with right in the beginning (see equation (\ref{ndf})). 
  
\subsection{Matter, loops and strings} 

The inclusion of fundamental matter fields is also natural in the prepotential 
formulation. It simply increases the number of intertwining operators at every 
lattice site. For simplicity, we introduce scalar matter field operators: 
$\left(\phi_{\alpha}(n), \phi^{*}_{\alpha}(n)\right)$ and their conjugate momenta 
$\left(\pi_{\alpha}(n), \pi^{*}_{\alpha}(n)\right)$ respectively. They are neutral 
under U(1) gauge transformations and transform as doublets under SU(2) gauge 
transformations:  $\phi_{\alpha}(n) 
\rightarrow \Lambda_{\alpha\beta}(n) \phi_{\beta}(n),~ \pi^{*}_{\alpha}(n) 
\rightarrow \Lambda_{\alpha\beta}(n) \pi^{*}_{\beta}(n)$. The canonical 
commutation relations are:
\bea 
\left[\pi_{\alpha}(n), \phi_{\beta}(m)\right] & = & 
\left[\pi_{\alpha}^{*}(n),\phi^{*}_{\alpha}(n)\right] 
~~=~  -i ~\delta_{\alpha\beta} \delta_{nm},  
\nonumber \\
\left[\pi_{\alpha}(n), \pi_{\beta}(m)\right] 
& = & \left[\phi_{\alpha}(n),\phi_{\beta}(m)\right] ~~ = ~~~~0, \\  
\left[\pi_{\alpha}(n), \pi^{*}_{\beta}(m)\right]
 &=&  \left[\phi_{\alpha}(n),\phi^{*}_{\beta}(m)\right]~~ = ~~~0.  \nonumber 
\label{ccrm} 
\eea 
The matter creation and annihilation operators are defined as: 
\bea 
{\sf a}^{\dagger}_{\alpha} (n) \equiv \frac{1}{\sqrt{2}}\left[\pi_{\alpha}(n) 
+ i \phi^{*}_{\alpha}(n)\right], &&   
{\sf a}_{\alpha} (n) \equiv \frac{1}{\sqrt{2}}\left[\pi^{*}_{\alpha}(n) 
- i \phi_{\alpha}(n)\right], 
\nonumber \\ 
{\sf b}^{\dagger}_{\alpha} (n) \equiv \frac{1}{\sqrt{2}}\left[\pi^{*}_{\alpha}(n) + 
i \phi_{\alpha}(n)\right], 
&&
{\sf b}_{\alpha} (n) \equiv \frac{1}{\sqrt{2}}\left[\pi_{\alpha}(n) - 
i \phi^{*}_{\alpha}(n)\right].    
\label{mcd} 
\eea
Like the prepotential operators they satisfy: 
\bea 
\left[{\sf a}_{\alpha} (n), {\sf a}^{\dagger}_{\beta} (m)\right] = 1,&&  
\left[{\sf b}_{\alpha} (n), {\sf b}^{\dagger}_{\beta} (m)\right] = 1, \nonumber \\
\Big[{\sf a}_{\alpha} (n), {\sf b}_{\beta} (m)\Big] = 0, &&
\left[{\sf a}_{\alpha} (n), {\sf b}^{\dagger}_{\beta} (m)\right] = 0.
\label{mco} 
\eea
Under SU(2): 
\bea
{\sf a}_{\alpha}(n) \rightarrow \Lambda_{\alpha\beta}(n) {\sf a}_{\beta}(n), ~~~~~~
{\sf b}^{\dagger}_{\alpha}(n) \rightarrow \Lambda_{\alpha\beta}(n) 
{\sf b}^{\dagger}_{\beta}(n), 
\label{gt4n}
\eea


\noindent they transform exactly like prepotentials (\ref{gt3n}) thus putting matter and 
gauge sector on the same footing under non-abelian gauge transformations. 
Therefore, we now have to  construct SU(2) singlets out of 
$(2d+2)$ types of SU(2) doublets of creation operators per lattice site instead 
of 2d types as in the case of pure gauge theory.  The orthonormal SU(2) gauge invariant 
states are now characterized by: 
\bea 
|j_1,j_2,j_{12},j_3,j_{123},........j_{12...(2d-1)},j_{2d},j_{12...2d}, j_{(2d+1)},
j_{12...(2d+1)} = j_{(2d+2)}  \rangle.  
\label{matter} 
\eea 
Further, the iterative method in appendix A again goes through and the states 
(\ref{matter}) can be easily constructed in terms of intertwining operators 
which will now  involve matter creation operators (\ref{mcd}) also. 

As the matter fields transform like SU(2) doublets and are neutral under U(1) 
gauge transformations, they provide end points for the abelian flux lines leading 
to a gauge invariant description of lattice gauge theories in terms of loops and 
strings.  Note that including scalar matter in the fundamental representation amounts to 
$d \rightarrow d+1$. The extra quantum numbers 
required per lattice site due to the inclusion of matter is $\left[4(d+1)-3\right] 
-[4d-3] =4$, as expected.  

\section{The dynamical issues} 

\noindent To discuss dynamics of loops, we consider pure SU(2) lattice gauge theory 
Hamiltonian \cite{kogut}:  
\bea 
H  =  \frac{g^{2}}{2} \sum_{n,i} E^2(n,i) 
+ \frac{2}{g^2}  \sum_{\square} \left(2-{\textrm Tr} U_{\square}\right) 
\label{ham}   
\eea 
where, 
\bea 
U_{\square} = U_{\rm plaquette} = U(n,i)U(n+i,j)U^{\dagger}(n+j,i)U^{\dagger}(n,j), 
\label{pl} 
\eea 
g is the coupling constant and $E^2(n,i) = E^2_L(n,i)= E^2_R(n+i,i)$ (see (\ref{consn})).
The loop states discussed in section (2) trivially diagonalize  the electric field term 
with eigenvalues $\sum_{loops} j(n,i)\left(j(n,i)+1\right)$ where 
$\sum_{loops}$ denotes summation over all the links on the loops.   
The electric field term in this loop basis is like  
potential energy term and counts the number of abelian flux lines on that 
link. The plaquette term acts  like kinetic energy term:  it 
makes the loops fluctuate over the corresponding plaquette by creating and 
destroying the abelian flux lines (\ref{dhn}). 

\subsection{The loop dynamics} 

\begin{figure}[b]
\begin{center}
\includegraphics[width=0.3\textwidth,height=0.28\textwidth]
{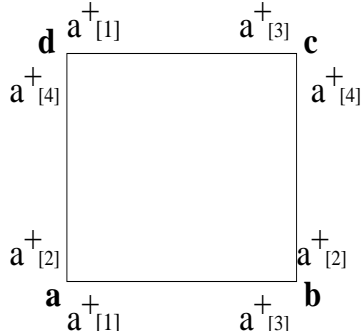} 
\end{center}
\vspace{-5mm}
\caption{The plaquette ${\bf abcd}$  with the corresponding prepotential operators from 
Figure (\ref{fig3}).}    
\label{fig:ABCD}
\end{figure}

As mentioned in the introduction, we restrict ourselves to d=2
and generalize the results to arbitrary d dimension at the end. 
We consider a plaquette abcd as shown in the Figure (\ref{fig:ABCD}) with the 
four edges $({\bf ab}), ({\bf bc}), ({\bf cd}), ({\bf da})$ denoted by $l_1,l_2,l_3,l_4$ 
respectively.  Using (\ref{dhn}) and (\ref{pari}), we write the gauge invariant plaquette 
operator over $abcd$ in terms of the prepotentials:   
\bea 
~~~~~ {\textrm Tr} U_{abcd}  = F_{abcd} \Big[\left(a^{\dagger}[1] \cdot \tilde{a}^{\dagger}[2]\right)_{a} 
\left (a^{\dagger}[2] \cdot\tilde{a}^{\dagger}[3]\right)_{b} 
\left (a^{\dagger}[3]\cdot\tilde{a}^{\dagger}[4]\right)_{c} 
\left (a^{\dagger}[4]\cdot\tilde{a}^{\dagger}[1]\right)_{d} 
 + \sum_{i=1}^{4} \pi(l_{i}) +~~~~~ 
\nonumber \\ 
 \sum_{i,j>i=1}^{4} \pi(l_{i}) \pi(l_{j})  
+ \sum_{i,j>i,k>j=1}^{4} \pi(l_{i}) \pi(l_{j}) \pi(l_{k}) 
+ \pi(l_{1})\pi(l_{2})\pi(l_{3})\pi(l_{4})\Big]F_{abcd}  \equiv \sum_{\alpha\beta\gamma\delta= \pm} 
H^{}_{\alpha\beta\gamma\delta} ~~   
\label{abcd} 
\eea

where $F_{abcd} \equiv F(l_1)F(l_2)F(l_3)F(l_4)$. 
In (\ref{abcd}), there are sixteen $SU(2)\otimes U(1)$ gauge invariant 
terms which are produced by substituting (\ref{dhn}) in (\ref{pl}). 
The first plaquette operator on the r.h.s. of the top equation in 
(\ref{abcd}) is written explicitly in terms of the 
prepotential intertwining operators at lattice sites {\bf a}, {\bf b}, {\bf c} and 
{\bf d} in Figure (\ref{fig:ABCD}). 
It  increases angular momenta 
by $\frac{1}{2}$ on the four links of $abcd$ and therefore we represent it 
by  $H^{}_{++++}$.  The rest of the 15 
plaquette operators in (\ref{abcd}) are generated by the action of $\pi$  on it.  
The single $\pi$ operation at any of the 4 links $(l_1,l_2,l_3,l_4)$ 
acting on $H^{}_{++++}$  produces 4 terms:~ $H^{}_{-+++}, H^{}_{+-++}, 
H^{}_{++-+}, H^{}_{+++-}$.  Similarly, the double $\pi$ operation 
produces 6 operators:~ $H^{}_{--++},H^{}_{-+-+},H^{}_{-++-},H^{}_{+--+},
H^{}_{+-+-},H^{}_{++--}$.  
There are four 3 $\pi$ terms in (\ref{abcd}): 
~$H^{}_{+---}, H^{}_{-+--}, H^{}_{--+-}, H^{}_{---+}$. 
Finally,  the 4 $\pi$ operation on all the four links produces 
a single term:  $H^{}_{----}$. 
Note that $\left(H^{}_{\alpha\beta\gamma\delta}\right)^{\dagger} = 
H^{}_{-\alpha-\beta-\gamma-\delta}$. The advantage of the 
form (\ref{abcd}) is that now this magnetic field term can be 
analyzed locally at {\bf a}, {\bf b}, {\bf c} and {\bf d} in the  
manifestly SU(2) gauge invariant way.    
To compute the action of $U_{abcd}$, it is convenient to 
label the loop state (\ref{std2}) at abcd as $|j_{abcd} \rangle$: 
\bea 
|j_{abcd} \rangle  \equiv |j_{1}^{a}j_{2}^{a}j_{3}^{a}j_{4}^{a}j_{12}^{a}  \rangle  \otimes 
|j_{1}^{b}j_{2}^{b}j_{3}^{b}j_{4}^{b}j_{12}^{b}  \rangle  \otimes 
|j_{1}^{c}j_{2}^{c}j_{3}^{c}j_{4}^{c}j_{12}^{c}  \rangle  \otimes 
|j_{1}^{d}j_{2}^{d}j_{3}^{d}j_{4}^{d}j_{12}^{d}  \rangle. 
\label{plst} 
\eea
\begin{figure}[t]
\begin{center}
\includegraphics[width=0.98\textwidth,height=0.42\textwidth]
{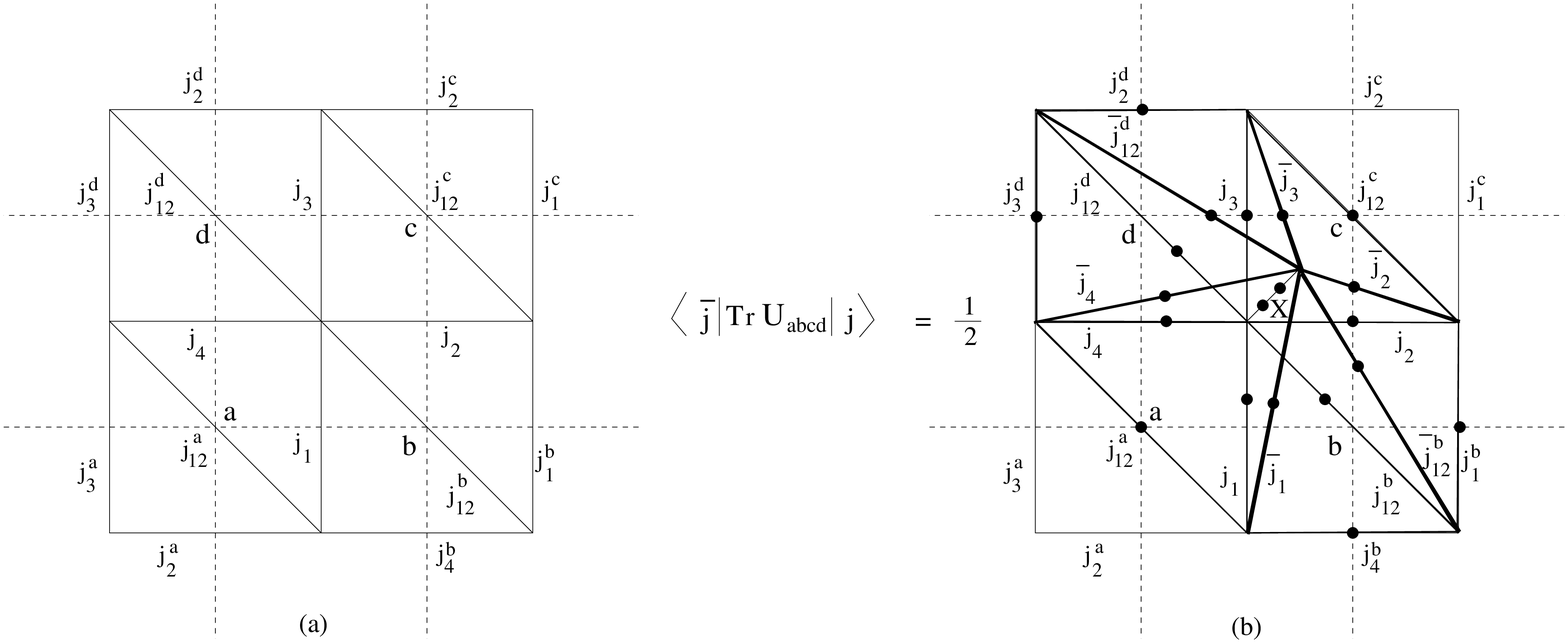} 
\end{center}
\vspace{-5mm}
\caption{From loop kinematics to loop dynamics. (a) The angular momenta satisfying 
SU(2) Gauss law on the dual lattice, (b) The matrix elements 
$\langle \bar{j}_{abcd}|{\textrm Tr} U_{abcd}| j_{abcd} 
\rangle$ with $X=\frac{1}{2}$ (also see \cite{sharat2}). The six tetrahedron  
are the six $6j$ symbols in (\ref{dyna1}). The unchanged j lines 
represent the delta function and a $\bullet$ over a j line   
represents the factor $(-1)^{j} \Pi(j)$ leading to ${\cal M}_{abcd}$ 
in (\ref{dyna1}).}     
\label{fig6}
\end{figure}
The U(1) Gauss law demands: 
\bea 
j^{a}_{1} = j^{b}_{3} \equiv j_1, ~j^{b}_{2} = j^{c}_{4} \equiv j_2,~  
j^{c}_{3} = j^{d}_{1} \equiv j_3, ~ j^{d}_{4} = j^{a}_{2} \equiv j_4. 
\label{u1gl} 
\eea
The matrix elements of ${\textrm Tr} U_{abcd}$ are computed directly  
in the loop basis (\ref{plst}) in appendix B using generalized Wigner-Eckart 
theorem and Biedenharn-Elliot identities. The final result is:   
\bea
\langle \bar{j}_{abcd}|{\textrm Tr}U_{abcd}| {j}_{abcd} \rangle  & = &  
{\cal M}_{abcd}  
{\left \{ \begin{array}{cccc}
{j}_{1} &  \bar{j}_{1} & \frac{1}{2}  \\
\bar{j}_{4} & {j}_{4}  & {j}^a_{12}   \\ 
\end{array} \right\}}   
{\left\{ \begin{array}{cccc}
{j}^b_{12} & \bar{j}^b_{12} & \frac{1}{2}  \\
\bar{j}_{1} & j_{1} & j_4^b\\
\end{array} \right \}}  
{\left\{ \begin{array}{cccc}
{j}^b_{12} & \bar{j}^b_{12} & \frac{1}{2}  \\
\bar{j}_{2} & j_{2} & j_1^b\\
\end{array} \right \}} 
\nonumber  \\ \nonumber  \\ &&
\hspace{1cm} {\left \{ \begin{array}{cccc}
{j}_{3} &  \bar{j}_{3} & \frac{1}{2}  \\
\bar{j}_{2} & {j}_{2}  & {j}^c_{12}   \\ 
\end{array} \right\}}   
{\left\{ \begin{array}{cccc}
{j}^d_{12} & \bar{j}^d_{12} & \frac{1}{2}  \\
\bar{j}_{3} & j_{3} & j_2^d\\
\end{array} \right \}}  
{\left\{ \begin{array}{cccc}
{j}^d_{12} & \bar{j}^d_{12} & \frac{1}{2}  \\
\bar{j}_{4} & j_{4} & j_3^d \\
\end{array}\right\}}. 
\label{dyna1} 
\eea
In (\ref{dyna1}), ${\cal M}_{abcd} \equiv D_{abcd} N_{abcd} P_{abcd}$ where: 
\bea 
D_{abcd} & = &  \delta_{j^a_{3},\bar{j}^a_{3}} 
\delta_{j^a_{4},\bar{j}^a_{4}} 
\delta_{j^a_{12},\bar{j}^a_{12}}  
\delta_{j_1^b,\bar{j}_1^b} \delta_{j_4^b,\bar{j}_4^b}  
\delta_{j^c_{1},\bar{j}^c_{1}} \delta_{j^c_{2},\bar{j}^c_{2}} \delta_{j^c_{12},\bar{j}^c_{12}} 
\delta_{j_2^d,\bar{j}_2^d} \delta_{j_3^d,\bar{j}_3^d}, 
\nonumber \\ 
\label{cnf} 
N_{abcd} & = & 
{\Pi}\left(j_1,\bar{j}_{1}, j_2,\bar{j_2},j_3,\bar{j_3},j_4,\bar{j_4},j^b_{12},\bar{j}^b_{12},
j^d_{12},\bar{j}^d_{12}\right) 
\\
P_{abcd} & = &  - (-1)^{j_1+j_2+j_1^b+j_4^b} (-1)^{j_3+j_4+j_3^d+j_2^d}  
\triangle(\bar{j}_1,\bar{j}_4,j_{12}^{a}) 
\triangle(\bar{j}_2,\bar{j}_3,j_{12}^{c}) 
\triangle(\bar{j}^b_{12},{j}^b_{12},\frac{1}{2})  
\triangle(\bar{j}^d_{12},{j}^d_{12},\frac{1}{2}).   
\nonumber 
\eea
In (\ref{cnf}), $D_{abcd}$ describes the trivial $\delta$ functions over the angular 
momenta which do not change under the action of the plaquette operator (\ref{abcd}), 
$N_{abcd}$ and $P_{abcd}$ give the corresponding numerical and the phase factors respectively. 
The multiplicity factors are:  $\Pi(x,y,...) \equiv \sqrt{(2x+1)(2y+1)...}$ and 
$\triangle(x,y,z)$ represent the phase factors associated with a triangle with sides x, y, z: 
$\triangle(x,y,z)  \equiv  (-1)^{x+y+z} \Rightarrow \triangle(x,y,z) =\pm 1$.   
The matrix elements (\ref{dyna1}) describe the dynamics in the loop basis 
(\ref{std2}) and can be geometrically represented by the Figure (\ref{fig6}b). 
This dynamics contains  three physical discrete angular 
momentum loop co-ordinates numbers per lattice site. The matrix elements (\ref{dyna1}) 
have been obtained\footnote{Our phase factors in (\ref{cnf}) are different resulting 
in real and symmetric matrix $\langle {\bar j}_{abcd} |{\rm Tr} U_{abcd} | j_{abcd} 
\rangle$ in (\ref{dyna1}).} 
in the context of dual description \cite{sharat2,robson} of  $(2+1)$ 
dimension lattice gauge theory in terms of  triangulated surfaces. 
It is easy to see that the matrix elements in (\ref{dyna1}) are symmetric: the 6j 
symbols satisfy  
${\left \{ \begin{array}{cccc}
x  &  \bar{x}  & u  \\
\bar{y}  &  y  & v   \\ 
\end{array} \right\}} = 
{\left \{ \begin{array}{cccc}
\bar{x}  &  {x}  & u  \\
{y}  &  \bar{y}  & v   \\ 
\end{array} \right\}}$  
and the factors $D_{abcd}, N_{abcd}$ and $P_{abcd}$ are individually symmetric under
$j_{abcd} \leftrightarrow \bar{j}_{abcd}$. The matrix elements in (\ref{dyna1}) 
are also real as  ${\rm Tr} U_{\rm plaquette}$ is a Hermitian operator. This reality 
can again be easily seen as the 6j symbols, $D_{abcd}, N_{abcd}, \triangle(abc)$ are 
themselves real. The remaining two phase factors $(-1)^{j_1+j_2+j_1^b+j_4^b}$ 
and $(-1)^{j_3+j_4+j_2^d+j_3^d}$ in $P_{abcd}$ are real because 
 $(j_1,j_2,j_1^b,j_4^b)$ and $(j_3,j_4,j_3^d,j_2^d)$ are the coordinates of the 
loop states at {\bf b} and {\bf d} respectively. Therefore, (\ref{part})  implies:
$j_1+j_2+j_1^b+j_4^b = {Integer}, ~ j_3+j_4+j_2^d+j_3^d = {Integer} \Rightarrow 
P_{abcd} = \pm 1$.  At this stage, before generalizing the loop dynamics 
to arbitrary d dimension, we cross check the d = 2 result (\ref{dyna1}).  
As the six 6j symbols and the $\delta$ functions in $D_{abcd}$ are geometrical in 
origin, we only need to check the numerical and the phase factors $N_{abcd},~ P_{abcd}$ 
respectively.  For this purpose, 
we replace ${\rm Tr} U_{abcd}$ by the identity operator ${\cal I}$. The computations in appendix B 
imply that now we only have to replace ${\frac{1}{2}}$ in each of the six 6j symbols in 
(\ref{dyna1}) and in $P_{abcd}$ in (\ref{cnf}) by 0.   
\begin{figure}[t]
\begin{center}
\includegraphics[width=0.4\textwidth,height=0.4\textwidth]
{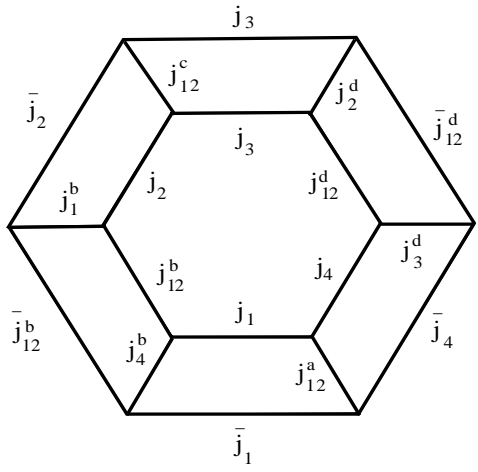} 
\end{center}
\vspace{-5mm}
\caption{ The $18j$ ribbon diagram representing exact SU(2) loop dynamics without any spurious 
gauge or loop degrees of freedom in d=2. The interior (exterior) edge carries the 
initial (final) angular momenta and the six bridges carry the angular momenta which 
are invariant under the action of ${\textrm Tr} U_{\square}$.  The six bridges 
along with the respective four angular momenta attached represent  
the six $6j$ symbols appearing  in (\ref{dyna3}) with $\bar{j}_i = j_i \pm \frac{1}{2}$.} 
\label{fig7}
\end{figure}
Using the value 
${\left \{ \begin{array}{cccc}
a  &  \bar{a}  & 0  \\
\bar{b}  &  b  & d   \\ 
\end{array} \right\}} 
= (-1)^{a+b+d} \delta_{a,\bar{a}} \delta_{b,\bar{b}} \left(\Pi(a,b)\right)^{-1}$, 
we get: 
\bea 
\begin{array}{cc}{\left \{ \begin{array}{cccc}
{j}_{1} &  \bar{j}_{1} & 0  \\
\bar{j}_{4} & {j}_{4}  & {j}^a_{12}   \\ 
\end{array} \right\}}   
{\left\{ \begin{array}{cccc}
{j}^b_{12} & \bar{j}^b_{12} & 0  \\
\bar{j}_{1} & j_{1} & j_4^b\\
\end{array} \right \}}  
{\left\{ \begin{array}{cccc}
{j}^b_{12} & \bar{j}^b_{12} & 0  \\
\bar{j}_{2} & j_{2} & j_1^b\\
\end{array} \right \}} 
\\  
{\left \{ \begin{array}{cccc}
{j}_{3} &  \bar{j}_{3} & 0  \\
\bar{j}_{2} & {j}_{2}  & {j}^c_{12}   \\ 
\end{array} \right\}}  
{\left\{ \begin{array}{cccc}
{j}^d_{12} & \bar{j}^d_{12} & 0  \\
\bar{j}_{3} & j_{3} & j_2^d\\
\end{array} \right \}}  
{\left\{ \begin{array}{cccc}
{j}^d_{12} & \bar{j}^d_{12} & 0  \\
\bar{j}_{4} & j_{4} & j_3^d \\
\end{array}\right\}}\end{array}   = 
\frac{\delta_{j_1,\bar{j}_1} 
\delta_{j_2,\bar{j}_2} 
\delta_{j_3,\bar{j}_3} 
\delta_{j_4,\bar{j}_4} 
\delta_{j^{b}_{12},\bar{j}^{b}_{12}} 
\delta_{j^{d}_{12},\bar{j}^{d}_{12}}}{N_{abcd}~P_{abcd}}, 
\label{new} 
\eea
Geometrically, the equation (\ref{new})  corresponds to putting 
$X = 0$ in the Figure (\ref{fig6}b). It implies   
$\langle \bar{j}_{abcd} |{\cal I}| {j}_{abcd} \rangle 
= \delta_{\bar{j}_{abcd},{j}_{abcd}}$ confirming 
the numerical and the phase factors in (\ref{dyna1}). 
We now write (\ref{dyna1}) in a more compact form which can be directly 
generalized to higher dimension. Henceforth, 
we ignore $D_{abcd}$ representing  trivial $\delta$ functions in 
(\ref{dyna1}). We write: 
\bea
&& \langle \bar{j}_{abcd}|{\textrm Tr}U_{abcd}| {j}_{abcd} \rangle   =   
N_{abcd}  \sum_{x} (2x+1) (-1)^{r+2x} 
{\left \{ \begin{array}{cccc}
{j}_{1} &  \bar{j}_{1} & x  \\ 
\bar{j}_{4} & {j}_{4}  & {j}^a_{12}   \\
\end{array} \right\}}   
{\left\{ \begin{array}{cccc}
\bar{j}_{4}&  j_{4}  & x  \\
{j}^d_{12} & \bar{j}^d_{12} & j_3^d \\
\end{array}\right\}} ~~ 
\nonumber \\ 
\nonumber \\ 
\nonumber \\ 
&& 
{\left\{ \begin{array}{cccc}
{j}^d_{12} & \bar{j}^d_{12} & x  \\
\bar{j}_{3} & j_{3} & j_2^d\\
\end{array} \right \}}  
{\left \{ \begin{array}{cccc}
\bar{j}_{3} &  {j}_{3} & x  \\
{j}_{2} & \bar{j}_{2}  & {j}^c_{12}   \\ 
\end{array} \right\}}   
{\left\{ \begin{array}{cccc}
j_{2}  &  \bar{j}_{2}  & x  \\
\bar{j}^b_{12} & {j}^b_{12} & j_1^b\\
\end{array} \right \}} 
{\left\{ \begin{array}{cccc}
\bar{j}^b_{12} & {j}^b_{12} & x  \\
{j}_{1} & \bar{j}_{1} & j_4^b\\
\end{array} \right \}} 
\prod_{i=1}^{4} 
\left(\delta_{\bar{j}_i,j_i + \frac{1}{2}} + \delta_{\bar{j}_i,j_i - \frac{1}{2}}\right) 
\nonumber \\
\nonumber \\ 
\nonumber \\ 
&&= N_{abcd}\underbrace{ \left[ \begin{array}{ccccccccccccc}
j_1   &    & j_4   &   &  {j}^d_{12}  &   &  j_3  &    &  j_2  &   &  {j}^b_{12}  &   \\
   & {j}^a_{12}    &    & j_3^d  &   & j_2^d  &    & {j}^c_{12}   &   & j_1^b  &   & j_4^b  \\
\bar{j}_1   &    &  \bar{j}_4  &   & \bar{j}^d_{12}  &     & \bar{j}_3  &    & \bar{j}_2   &   &  \bar{j}^b_{12}  &    \\
\end{array} \right]}_{\rm 18j~ coefficient~ of~ the ~ second~ kind} 
\prod_{i=1}^{4} 
\left(\delta_{\bar{j}_i,j_i + \frac{1}{2}} + \delta_{\bar{j}_i,j_i - \frac{1}{2}}\right) 
\label{dyna3} 
\eea
The 18j symbols in (\ref{dyna3}) are shown in (\ref{fig7}). 
Note that  $P_{abcd} \left(= (-1)^{r+1} \right)$ in (\ref{cnf}) is precisely the phase factor 
needed to define 18j symbol \cite{yut} in (\ref{dyna3}). 
Further, the 12 triangular constraints in (\ref{dyna3}) at the 12 vertices of the 
ribbon diagram in Figure (\ref{fig7}) are already solved in terms of the 
linking numbers. Therefore it is only the value of the $3nj = 18j$ (n = 6) symbol which 
is important. The form (\ref{dyna3}) also makes reality and symmetry of 
$\langle \bar{j}_{abcd} | {\rm Tr U_{abcd}} | j_{abcd} \rangle$ manifest as 
3nj symbols of second kind are real and symmetric: 
\bea  
&& \left[ \begin{array}{ccccccccccccc}
j_1   &    & j_4   &   &  {j}^d_{12}  &   &  j_3  &    &  j_2  &   &  {j}^b_{12}  &   \\
   & {j}^a_{12}    &    & j_3^d  &   & j_2^d  &    & {j}^c_{12}   &   & j_1^b  &   & j_4^b  \\
\bar{j}_1   &    &  \bar{j}_4  &   & \bar{j}^d_{12}  &     & \bar{j}_3  &    & \bar{j}_2   &   &  \bar{j}^b_{12}  &    \\
\end{array} \right]  \nonumber \\ \nonumber \\
= && \left[ \begin{array}{ccccccccccccc}
\bar{j}_1   &    & \bar{j}_4   &   &  \bar{j}^d_{12}  &   &  \bar{j}_3  &    
&  \bar{j}_2  &   &  \bar{j}^b_{12}  &   \\
  & {j}^a_{12}  &  & j_3^d  &  & j_2^d  &  & {j}^c_{12}  &  & j_1^b  &  & j_4^b  \\
{j}_1 & &  {j}_4  &  & {j}^d_{12}  &   & {j}_3  &  & {j}_2  &   &  {j}^b_{12}  &    \\
\end{array} \right]  \nonumber 
\eea 

Before going to arbitrary dimension, we make the following simple 
observation. Let $\Delta N_{x},~ x=a,b,c,d$ denote the number of angular momenta appearing 
in the loop states $|j_{abcd}\rangle$ in (\ref{plst})  
which change under the action of the plaquette operator 
${\textrm Tr} U_{abcd}$  at lattice site x. 
In the present, $d=2$, case: 
\bea 
\Delta N_{a} = 2,~\left(j_1^a,j_2^a\right);~~ \Delta N_b= 3, ~ \left(j_2^b,j_3^b,j_{12}^b\right); ~~ 
\Delta N_c=2, ~ \left(j_3^c,j_4^c\right); ~~\Delta N_d = 3, ~ \left(j_1^d,j_4^d,j_{12}^{d}\right). 
\nonumber  
\eea
The $U(1)$ identification (\ref{u1gl})  
implies double counting on each of the 4 links of the plaquette $abcd$.  
Therefore, the number of angular momenta which change under the action of the 
plaquette in the $(12)$ plane: $\Delta N(12) = \Delta N_a+\Delta N_b
+\Delta N_c+\Delta N_d-4 = 10-4 = 6 = n$. 
This analysis  will be useful to generalize the loop dynamics 
to arbitrary dimensions below.

\subsection{d dimension} 

\begin{figure}[t]
\begin{center}
\includegraphics[width=0.95\textwidth,height=0.32\textwidth]
{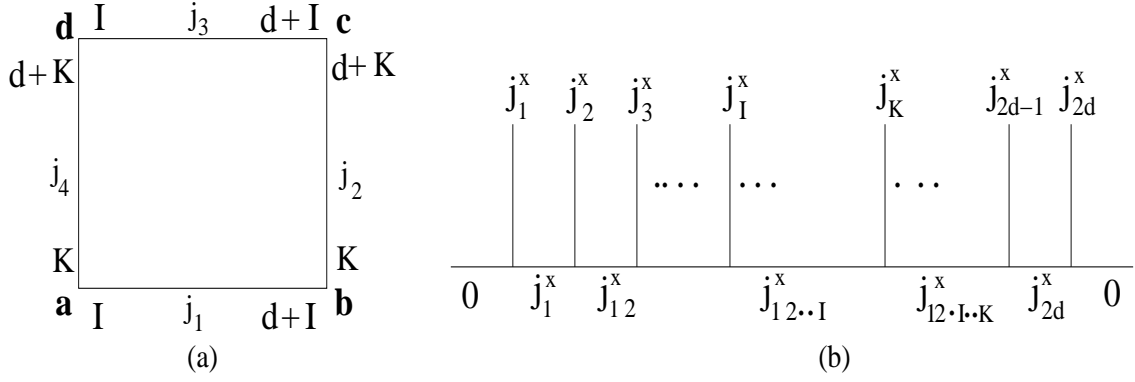} 
\end{center}
\vspace{-5mm}
\caption{(a) The plaquette ${\bf abcd}$ in the $(I,K)$ plane in d dimension. We choose 
$I < K,~ 1 \le I < d$ and  $1 < K \le d$, (b) The angular momentum addition scheme at 
site x (={\bf a},{\bf b},{\bf c},{\bf d}). Note that $j_1^x$ and $j_{2d}^x$ appear 
twice in the scheme. The $\delta$ functions are subtracted in (\ref{nwc}) to avoid this 
double counting.} 
\label{fig8}
\end{figure}
\begin{figure}[t]
\begin{center}
\includegraphics[width=1.0\textwidth,height=0.315\textwidth]
{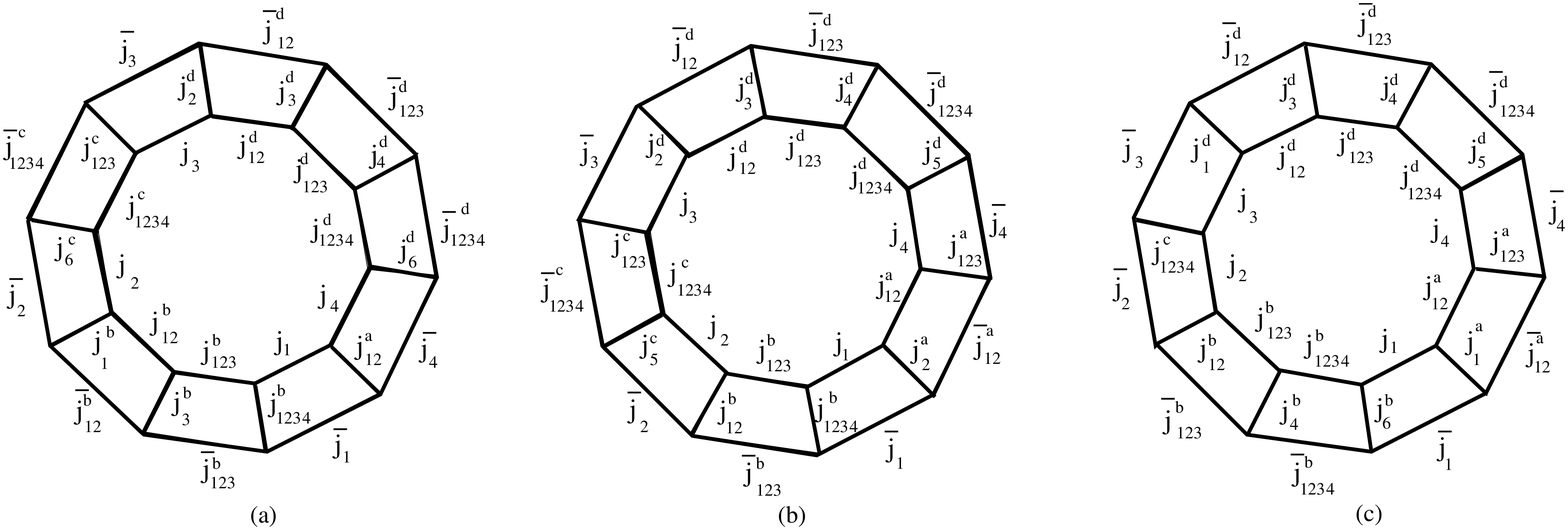} 
\end{center}
\vspace{-5mm}
\caption{The $30j$ ribbon diagrams representing exact SU(2) loop dynamics without 
any spurious gauge or loop degrees of freedom in d=3: a) (12) plane, b) (13) plane, 
c) (23) plane. The angular momenta $j_1, j_2, j_3$ and $j_4$ are as shown in Figure 
(\ref{fig8}a). 
The initial (inner) and final (outer) angular momenta differ by $\frac{1}{2}$.} 
\label{fig9}
\end{figure}

It is clear from the previous section that the loop dynamics in $d$ dimension 
is also given in terms of $3nj$ symbols. However, in arbitrary d dimension, 
n will depend on the orientation of the plaquette. We will now compute n. 
We consider the plaquette {\bf abcd} in the $(I,K), I < K$ plane as shown in 
Figure (\ref{fig8}). Like in $d=2$, we consider the 
loop states over the plaquette $abcd$: 
\bea 
&& |j_{abcd}\rangle \equiv |LS\rangle_a \otimes
|LS\rangle_b \otimes|LS\rangle_c \otimes|LS\rangle_d 
\label{x1y1} 
\eea  
where $|LS\rangle_{x=a,b,c,d} = |j^x_1,j^x_2,j^x_{12},j^x_3,j^x_{123},....
,j^x_{I},j^x_{12...I},..,j^x_{K},j^x_{12..I..K},....
,j^x_{2d-1},j^x_{12..(2d-1)}(=j^x_{2d}),j^x_{2d},0 \rangle$.  
We now have to count the the number of angular momenta in (\ref{x1y1}) 
which  change under the action of the plaquette operator ${\textrm Tr} U_{abcd}$ in 
the $(IK)$ plane. With the choice $I<K~ (1 \le I < d, 1 < K \le d)$, we have: 
\bea 
\Delta N_a= 2+(K-I)-\delta_{I,1},  &~~ & \Delta N_b= 2+(d+I)-K,  \nonumber \\
\Delta N_c = 2 + (K-I) - \delta_{K,d},  &~~ & \Delta N_d= 2+ (d +K) -I
-\delta_{I,1} -\delta_{K,d} 
\label{nwc}
\eea
This implies:
\bea 
\Delta N(IK) = \Delta N_a+\Delta N_b+\Delta N_c+\Delta N_d -4 = 
2\left[2+d+(K-I)-\delta_{I,1}-\delta_{K,d}\right] = n(IK).  
\label{numbb}
\eea
Like in d =2 case, we have subtracted 4 in (\ref{numbb}) because of U(1) gauge invariance. 
Note that for d=2, $\Delta N(12)  =6$ and for d=3, $\Delta N(12)= 
\Delta N(13) = \Delta N(23) = 10$.  The d = 3 loop dynamics is explicitly shown in Figure 
(\ref{fig9}) where we have used the notations from Figure (\ref{fig8}a), i.e.: 
\bea 
(I=1, K=2) = (12)~{\rm plane}: j_1^a=j_4^b=j_1,~j_2^b=j_5^c=j_2,~j_4^c=j_1^d=j_3,j_5^d=j_2^a=j_4, \nonumber \\
(I=1, K=3) = (13)~{\rm plane}: j_1^a=j_4^b=j_1,~j_3^b=j_6^c=j_2,~j_4^c=j_1^d=j_3,j_6^d=j_3^a=j_4, \nonumber \\
(I=2, K=3) = (23)~{\rm plane}: j_2^a=j_5^b=j_1,~j_3^b=j_6^c=j_2,~j_5^c=j_2^d=j_3,j_6^d=j_3^a=j_4. \nonumber 
\eea 
It is clear from (\ref{numbb}) 
that in  higher $(d > 3)$ dimension $\Delta N(IK)$ depends on the orientation of the 
plaquette.  The corresponding $3n(I,K)j$ symbol describing the loop dynamics in the 
above angular momentum addition scheme can be easily written down.

\section{SU(N) prepotentials} 

We now briefly discuss the extension of the ideas in this paper to the 
SU(N) gauge group. The SU(N) generalization of the SU(2) Jordan-Schwinger 
mapping (\ref{sbn}) has been done in \cite{manu3} in the context of SU(N) 
coherent states. We define the left and right SU(N) electric fields
through the SU(N) prepotentials:  
\bea
E^{a}_{L} = \sum_{r=1}^{N-1} E^{a}_{L}[r] = \sum_{r=1}^{N-1} 
a^{\dagger}[r] \frac{\lambda^{a}[r]}{2} a[r]; ~~~
E^{a}_{R} = \sum_{r=1}^{N-1}E^{a}_{R}[r] = \sum_{r=1}^{N-1} 
b^{\dagger}[r]\frac{\lambda^{a}[r]}{2} b[r].
\label{sbnn}
\eea
In (\ref{sbnn}), the index  $r(=1,2,...,(N-1))$ covers all the $(N-1)$ fundamental 
representations of SU(N) group and all the link indices (n,i) are suppressed. Again, 
the $(N - 1)$ SU(N) Casimirs on the left 
and the right ends of any link (n,i) are the prepotential number operators: 
\bea 
N_{a}[r] = a^{\dagger}[r] \cdot a[r], ~~~~ N_{b}[r] = b^{\dagger}[r] \cdot b[r]; ~~
r=1,2,..,(N-1).  
\label{sun1} 
\eea
The SU(N) kinematical constraints involving  SU(N) Casimirs \cite{sharat1} are now in 
terms of the prepotential number operators: 
\bea 
N_{a}[r](n,i) = N_{b}[r](n+i,i), ~~ r=1,2,..(N-1). 
\label{sun2}
\eea
The $(N-1)$ constraints (\ref{sun2}) are analogous to the single constraint 
(\ref{noe}) for SU(2).  
The defining relations (\ref{sbnn}) for the prepotentials 
and the constraints (\ref{sun2}) imply the following abelian 
gauge invariance:
\bea
a^{\dagger}[r](n,i) \rightarrow expi\theta[r](n,i)~ a^{\dagger}[r](n,i); ~~~~
b^{\dagger}[r](n+i,i) \rightarrow exp - i\theta[r](n,i) ~b^{\dagger}[r](n+i,i).
\label{u1n}
\eea
Thus the SU(N) prepotential formulation will have $SU(N) \otimes \left(U(1)\right)^{N-1}$ 
gauge invariance leading to $(N-1)$ varieties of loops. Therefore, using the ideas of 
SU(N) Schwinger bosons and the corresponding coherent states \cite{manu3}, like in 
appendix A for SU(2), 
it should be possible to find an orthonormal loop basis and corresponding loop dynamics 
for SU(N) lattice gauge theory
as well. The work in this direction is in progress.

\section{Summary and discussion} 

In this work we developed ideas and techniques to formulate  Hamiltonian lattice gauge 
theories exactly and most economically in terms of loop and string degrees of freedom. 
This required systematically solving Gauss law constraints, Mandelstam constraints and 
finally triangular constraints.  The apparently highly non-local and formidable Mandelstam 
constraints in terms of the link operators were cast and then solved locally in terms of the 
prepotential intertwining operators. Infact, one of the motivations for  this work 
was to develop manifestly SU(2) gauge invariant techniques involving 
gauge invariant local intertwining prepotential operators and 
intertwining/linking quantum numbers  having direct geometrical interpretation in terms of 
loops. The resulting simplifications have been emphasized in the text. 
Note that the loop states as well as loop dynamics were constructed directly 
in terms of the above, without using Clebsch-Gordan coefficients which 
are not gauge invariant and also do not have direct interpretation in terms of 
loops.  The final loop dynamics, i.e.,  the matrix elements of the magnetic field terms 
in between the two loop basis states, are found to be real and symmetric 
and are given by $3nj$ symbol in arbitrary dimension.  
Therefore, this loop space description of gauge theories is also a 
non-abelian dual description \cite{sharat2} where the effect of compactness 
of the gauge group is contained in the integer intertwining  or half-integer 
angular momentum quantum numbers labeling the loop states. In the simpler context of compact 
(2+1) and (3+1) U(1) gauge theories such duality transformations are known 
to isolate the topological magnetic monopole degrees of freedom leading to 
confinement \cite{banks}.  
It will also be interesting to develop a systematic 
weak coupling ($g \rightarrow 0$) loop perturbation theory near the continuum. 
This perturbation 
theory should encapsulate the global gauge invariant loop fluctuations as opposed to the 
local gauge field fluctuations which is the case with the standard perturbation theory. 
The issue of color confinement and vacuum structure will be of special interest. 
The work in this direction is in progress and will be reported 
elsewhere. The present prepotential approach has also been found useful  
to study spin networks and analyze the spectrum of the volume 
operators in lattice loop quantum gravity \cite{manu4}. 
Finally, the maximally reduced loop basis and the corresponding 
matrix elements should also be useful for numerical diagonalization.   
 
\begin{center} 
{\it This work is dedicated to the memory of late Prof. C. K. Majumdar.} 
\end{center} 

\vspace{1.00cm} 

\noindent{\bf \large Appendix A} 

\vspace{0.51cm} 

\noindent In this section we explicitly construct all possible   
orthonormal loop states (\ref{std2}) in terms of prepotentials 
intertwining operators.  We appropriately interpret, modify and 
generalize  the techniques developed in \cite{schwinger} to 
add angular momenta in terms of Schwinger bosons or 
equivalently prepotentials in our formulation. The basic idea is that angular 
momenta can be combined directly in terms of prepotentials and the $d(2d-1)$ 
intertwining operators (\ref{int}) by taking certain direct products of SU(2) 
coherent states. We explain this idea in $d=2$. It's generalization to 
arbitrary dimension is then obvious and done next. 
The SU(2) group manifold $S^3$ is characterized by a doublet of complex 
numbers $(z_1,z_2)$ with the constraint: $|z_1|^2 + |z_2|^2 =1$.  The  SU(2) 
coherent state in the spin j representation are  given by 
\cite{manu3}: 
\bea  
|z \rangle_{j}  =  
\sum_{m=-j}^{j} \frac{(z_1)^{j+m}(z_{2})^{j-m}}  
{\sqrt{(j+m)!(j-m)!}}~ |j,m \rangle  \equiv  
\sum_{m=-j}^{j} \phi_{jm}(z)~ |j,m \rangle  
\label{su2cs} 
\eea
In terms of Schwinger bosons, 
\bea 
|j,m\rangle \equiv \frac{(a^{\dagger}_1)^{j+m}(a^{\dagger}_{2})^{j-m}} 
{\sqrt{(j+m)!(j-m)!}}~ |0,0 \rangle. 
\label{sjm} 
\eea
The generating function of SU(2) coherent state is: 
\bea 
\sum_{j} \Phi_j(\delta) |z\rangle_{j}  =  exp \left( \delta~z \cdot a^{\dagger}\right) | 0 \rangle    
\equiv exp \left(\delta~(z_1 a_1^{\dagger}+ z_2 a_2^{\dagger}) \right)| 0 \rangle  
\label{csd}
\eea   
where  $\Phi_{j}(\delta) = \delta^{2j}$. The states $|j,m\rangle$ can be extracted 
by  comparing the terms with coefficients $\delta^{2j}$ on both sides of 
(\ref{csd}).  To add the 2 angular momenta, $J^a[1] = a^{\dagger}[1] \frac{\sigma^a}{2} a[1]$ 
and $J^a[2] = a^{\dagger}[2]\frac{\sigma^a}{2} a[2]$,  we consider 
direct product of the generating functions of two SU(2) coherent states 
defined over the complex planes  $(x_1,x_2)$ and $(y_1,y_2)$ respectively. 
\bea 
\vert x \rangle \otimes \vert y \rangle \equiv  
\sum_{j_{1}j_{2}} 
\vert x  \rangle_{j_{1}} \otimes ~\vert y \rangle_{j_{2}} 
= exp\left(x \cdot a^{\dagger}[1] + y \cdot a^{\dagger}[2]\right) \vert 0 \rangle  
\label{id} 
\eea
We apply the differential operator involving a triplet of complex parameters 
$(\delta_1,\delta_2,\delta_3)$ and a complex doublet  $z (\equiv (z_{1},z_{2}))$: 
\bea 
exp\left(\delta_{3}~(\partial_{x} \cdot \tilde{\partial}_{y})  + \delta_{1}~(z \cdot \partial_{x}) + 
\delta_{2}~ (z \cdot \partial_{y}) \right) 
\label{si} 
\eea  
on the both sides of (\ref{id}) and put $x = y =0$ to get \cite{schwinger}:
\bea
\sum_{j_{1}j_{2}j_{12}}\Phi_{j_1 j_2 j_{12}} (\vec{\delta})
\vert z \rangle_{j_{12}}^{j_1j_2}   
 =  exp\Big(\delta_{3}~a^{\dagger}[1] \cdot \tilde{a}^{\dagger}[2] + z \cdot 
a^{\dagger}[12]\Big)
=  exp\Big(\delta_{3}~ L_{12}  + z \cdot a^{\dagger}[12]\Big)
\vert 0 \rangle  
\label{12} 
\eea 
where $L_{12}$ is the intertwining operator in the (12) plane and
\bea 
a^{\dagger}_{\alpha}[12]   \equiv   \delta_{1} a^{\dagger}_{\alpha}[1] + 
\delta_{2} a^{\dagger}_{\alpha}[2] 
\label{idj}
\eea
\bea 
\Phi_{j_1 j_2 j_{12}} (\vec{\delta})  =  \left[\frac{(j_1 + j_2 + j_{12}+1)!}  
{(2j_{12}+1)}\right]^{\frac{1}{2}}   
\frac{(\delta_{1})^{j_1 - j_2 + j_{12}} (\delta_{2})^{-j_1 + j_2 + j_{12}} 
(\delta_{3})^{j_1 + j_2 - j_{12}}}
{\left[(j_1 - j_2 + j_{12})! (-j_1 + j_2 + j_{12})! 
(j_1 + j_2 - j_{12})!\right]^{\frac{1}{2}}}
\label{nnaa1}  
\eea 
In (\ref{12}), $|z \rangle_{j_{12}}^{j_1j_2}$ is the 
coherent state in the $j_{12}$ representation 
of the combined angular momentum $(\vec{J}[1]+\vec{J}[2])$, i.e: 
\bea 
\vert z \rangle_{j_{12}}^{j_1j_2} \equiv \sum_{m_{12}=-j_{12}}^{+j_{12}} \phi_{j_{12}m_{12}}(z)
\vert j_1 j_2 j_{12} m_{12} \rangle 
\label{cs12}
\eea
Note that $(j_1-j_2+j_{12}), (-j_1+j_2+j_{12})$ and $(j_1+j_2-j_{12})$ are all 
non-negative  integers due to angular momentum addition rules. 
The coherent state generating function (\ref{12})  in the representation $j_{12} 
~ (|j_1-j_2| \le j_{12} \le j_1+j_2)$ is the generalization of the generating 
function (\ref{csd}) in the case of single angular momentum. 
To illustrate this,  we put $\delta_2 =\delta_3 = 0 (\Rightarrow  j_2=0, j_1=j_{12})$ 
in (\ref{12}) to  recover (\ref{csd}). 
Now to project out $J_{total} = J[1] +J[2] =0$ (``gauge invariant state in d = 1") state  
from the left hand side of (\ref{12}), we put: 
\bea 
\delta_1 = 0, ~~\delta_2 = 0~~ \Rightarrow  ~~ j_{1} = j_2 \equiv j, ~~ j_{total} = 
j_{12}=0. \nonumber 
\eea 
Now (\ref{12}) takes the form: 
\bea 
|j_1=j,j_2=j,j_{total}=j_{12}=0 \rangle  = {\cal{N}}(j) 
\frac{\left(L_{12}\right)^{l_{12}}}{(l_{12})!} ~|0 \rangle  
\label{deq1}
\eea
In (\ref{deq1}) $l_{12}=2j$ and  
\bea 
{\cal{N}}(j) = {N(j_1=j,j_2=j,j_{12}=0)} = \frac{1}{\sqrt{(2j+1)}}  
\label{ncd1} 
\eea
In (\ref{ncd1}), $N(j_1,j_2,j_{12}) = \sqrt{\frac{{(2j_{12}+1)}}{(j_1+j_2+j_{12}+1)!}}
\left[(-j_1+j_2+j_{12})!(j_1-j_2+j_{12})!(j_1+j_2-j_{12})!\right]^{\frac{1}{2}}$
are the normalization constant so that 
$\langle j_1=j,j_2=j,j_{total}=j_{12}=0|j_1=j,j_2=j,j_{total}=j_{12}=0 \rangle  =1$.  
The U(1) Gauss law makes j site independent. These states are the 
trivial examples of the ``loop states" in d=1. 
We now generalize the above simple construction in d=1 to d=2. To add the 
third angular momentum corresponding to $a^{\dagger}[n,3]$ we 
rewrite (\ref{12}) with $z \rightarrow x$ and take the direct product with 
\bea 
\vert y \rangle \equiv  \sum_{j_{3}m_{3}} \phi_{j_{3}m_{3}}(y) 
\vert j_{3}, m_{3} \rangle = exp (y.a^{\dagger}[3]) \vert 0 \rangle \nonumber   
\eea
We now apply the operator (\ref{si}) with $\delta \rightarrow \sigma$: 
\bea 
exp\left(\sigma_{3}~(\partial_{x} \cdot \tilde{\partial}_{y})+\sigma_{1}~(z \cdot \partial_{x}) + 
\sigma_{2}~ (z \cdot \partial_{y}) \right) 
\label{sii} 
\eea  
to get: 
\bea
\sum_{j_{1}j_{2}j_{12}j_3j_{123}} \Phi_{j_1 j_2 j_{12}} (\vec{\delta}) 
\Phi_{j_{12} j_3 j_{123}} (\vec{\sigma}) |z \rangle^{j_1j_2j_{12}j_3}_{j_{123}}  
= exp \left({L}_{123} +  z.a^{\dagger}[123]\right)~ \vert 0 \rangle
\label{122} 
\eea 
where, 
\bea 
a^{\dagger}[123] \equiv \sigma_1 a^{\dagger}[12] + 
\sigma_{2} a^{\dagger}[3]  \hspace{2cm}  \nonumber \\  
{L}_{123}    \equiv  \delta_{3}~a^{\dagger}[1].\tilde{a}^{\dagger}[2] + 
\sigma_{3}~a^{\dagger}[12].\tilde{a}^{\dagger}[3]
 =  \delta_3 L_{12} + \sigma_3\left(\delta_1L_{13} +\delta_2L_{23}\right).  
\label{def}
\eea 
and 
\bea 
|z \rangle^{j_1j_2j_3}_{j_{12}j_{123}} & \equiv & 
\sum_{m_{123}=-j_{123}}^{+j_{123}} \phi_{j_{123}m_{123}}(z)
\vert j_1 j_2 j_{12} j_3 j_{123} m_{123} \rangle \nonumber 
\label{cs3} 
\eea
is the SU(2) coherent states in the combined angular momentum $\vec{J}[123]= 
\left(\left(\vec{J}[1]+\vec{J}[2]\right)+\vec{J}[3]\right)$ representation with 
$|j_{12}-j_3| \le j_{123} \le j_{12}+j_3$. The operator ${L}_{123}$ contains 
the intertwining operators in the (12), (13) and (23) planes. Note that, like in the previous case, 
putting $ \sigma_2=\sigma_3 =0 (\Rightarrow j_3=0, j_{123}= j_{12})$ in (\ref{def}) 
we recover (\ref{12}). Therefore, it is a sequential process. Repeating the 
same steps once again in the $4^{th}$ direction with the 
prepotential operators $a^{\dagger}[4]$, we finally get: 
\bea
\sum_{j_{1}j_{2}j_{12}j_3j_{123}j_{1234}} \Phi_{j_1 j_2 j_{12}} (\vec{\delta}) 
\Phi_{j_{12} j_3 j_{123}} (\vec{\sigma}) \Phi_{j_{123} j_4 j_{1234}}(\vec{\rho})
|z \rangle^{j_1j_2j_{12}j_3j_{123}j_4}_{j_{1234}}  
 =  exp~\left({L}_{1234} + z.a^{\dagger}[1234]\right)~  \vert 0 \rangle  
\label{123} 
\eea 
where, 
$a^{\dagger}[1234] \equiv  \rho_{1} a^{\dagger}[123] + \rho_{2} a^{\dagger}[4]$ and  
${L}_{1234}$ contains intertwining in (12), (13), (14), (23), (24) and (34) planes: 
\bea 
{L}_{1234} & \equiv & \delta_{3}~a^{\dagger}[1] \cdot\tilde{a}^{\dagger}[2] + 
\sigma_{3}~a^{\dagger}[12] \cdot\tilde{a}^{\dagger}[3]+ 
\rho_3 a^{\dagger}[123] \cdot\tilde{a}^{\dagger}[4] \nonumber \\ 
& = & \delta_3\left(L_{12}\right) +\sigma_3\left(\delta_1L_{13} + \delta_2 L_{23} \right) + 
\rho_3\left(\sigma_1\delta_1 L_{14} + \sigma_1\delta_2 L_{24} + \sigma_2 L_{34}\right) 
\eea
The gauge invariant states can now be projected (like in d=1 case) 
by choosing: 
\bea 
\vec{\rho} = (0,0,\rho_3)~~  => ~~  j_{123} = j_4, ~~ j_{1234}=0. \nonumber 
\eea 
We now compare the coefficients of $(\delta_1,\delta_2,\delta_3)$, $(\sigma_1,
\sigma_2,\sigma_3)$ and $\rho_{3}$ to get the all possible manifestly gauge 
invariant orthonormal states at site n.
After some algebra we get\footnote{Note that 
we have solved the SU(2) Gauss law at a site n and have ignored the 
site index from $j^{s}$ and $l^{s}$.}:  
\bea 
|j_1,j_2,j_{12} \rangle & \equiv & |j_{1},j_{2},j_{12},(j_3),j_{123}=
(j_{4}), j_4, j_{total} = j_{1234}=0, m_{total}=  m_{1234}=0 \rangle 
\nonumber    \\
&=& {\cal{N}}(j) \sum_{{}^{l_{13},l_{14}}_{l_{23},l_{24}}} 
\hspace{-0.05cm}{}^{{}^\prime} 
\prod_{{}^{i,j}_{i < j}} \frac{\big(L_{ij}(n)\big)^{l_{ij}}}{(l_{ij})!} | 0 \rangle 
\label{fr}
\eea
In (\ref{fr}), ${\cal{N}}(j)$ is the constant given in terms of $N(j_1,j_2,j_{12})$ 
in (\ref{nnaa1}): 
\bea 
{\cal{N}}(j) = {N(j_1,j_2,j_{12})}~{N(j_{12},j_3,j_{4})}~{N(j_4,j_4,0)}  
\label{norm1}
\eea
so that the states in (\ref{fr}) are normalized to unity. 
The d=2 results (\ref{fr}) and (\ref{norm1}) are analogous 
to d=1 results given in (\ref{deq1}) and (\ref{ncd1}) respectively.  
The difference is the summation over the linking numbers 
in (\ref{fr}) which was missing in (\ref{deq1}). This is because now 
there are many possibilities of linking or contracting the 4 types of 
Young boxes representing $a^{\dagger}[1], a^{\dagger}[2], a^{\dagger}[3] $ 
and $a^{\dagger}[4]$ mutually to produce $j_{total}=0$ states. 
These contractions must be such that the states (\ref{fr}) are 
the eigenstates of the CSCO(I) as well as  CSCO(II) listed 
in (\ref{amm2}). In the construction (\ref{fr}) the 
linking numbers have to satisfy: 
\bea 
\sum_{j=1}^{2d} l_{ij} = 2j_{i}, ~~ i =1,2,..,2d. 
\label{network0}  
\eea  
and 
\bea 
l_{12} &=& j_{1}+j_{2}-j_{12} \nonumber \\
\label{network1} 
l_{13}+l_{23} &=& j_{12}+j_{3}-j_{123} = j_{12}+j_{3}-j_{4} \\
l_{14}+l_{24} +l_{34} &=& j_{123}+j_{4}-j_{1234} = 2j_{4}. \nonumber 
\eea  
ensuring that the states are common eigenstates of 
CSCO(I) and  CSCO(II) respectively. 
The constraints (\ref{network0}) and (\ref{network1}) on the linking numbers are 
denoted by the $\prime$ over the summation sign in (\ref{fr}). 
The last constraint in (\ref{network1}) represents the partition 
of $j_{4}$ in terms of the linking numbers and is already contained in (\ref{network0}).  
Note that we have combined angular momenta directly in terms of $l_{ij}$ 
without using any Clebsch-Gordon coefficients containing magnetic quantum numbers. 
The generalization of these techniques to arbitrary dimension is a sequential 
process and the results are given in (\ref{std2}). 
 
\vspace{1cm} 

\noindent{\bf \large Appendix B} 

\vspace{0.51cm} 

\noindent In this appendix, we compute the SU(2) loop dynamics. 
As mentioned before the electric field term in (\ref{ham}) simply 
counts the abelian flux lines without changing the loop states. 
The action of the magnetic field term in (\ref{ham}) on the loop states is 
non-trivial. The calculations of the matrix elements of ${\rm Tr} U_{\square}$ 
in d=2 are  presented below. It is convenient to define the square root of the 
multiplicity factors: 
\bea 
\Pi({x,y,....}) \equiv \sqrt{(2x+1)(2y+1)....}, \nonumber 
\eea 
which will occur repeatedly below. 

\noindent The matrix elements of an intertwining 
operators $L_{12} = \left(a^{\dagger}[1] \cdot \tilde{a}^{\dagger}[2]\right)$ 
in the corresponding angular momentum basis $|j_1,j_2,j_{12},m_{12}\rangle$ 
are given by the generalized Wigner Eckart theorem \cite{varsha,yut}: 
\bea 
~~\langle \bar{j}_1,\bar{j}_2,\bar{j}_{12}, \bar{m}_{12}|
\left(a^{\dagger}[1] \cdot\tilde{a}^{\dagger}[2]\right)|j_1,j_2,j_{12},m_{12} \rangle  
 = \sqrt{2} ~ \langle \bar{j}_1,\bar{j}_2,\bar{j}_{12}, \bar{m}_{12}| 
\left(a^{\dagger}[1] \otimes {a}^{\dagger}[2]\right)^{0}_{0}
|j_1,j_2,j_{12},m_{12} \rangle ~~ \nonumber \\
= {\sqrt{2}}~ (-1)^{(\bar{j}_{12}-\bar{m}_{12})} 
~\Pi({j_{12},\bar{j}_{12},0})~ 
{\left( \begin{array}{cccc}
\bar{j}_{12} & 0 & {j}_{12}   \nonumber \\
- \bar{m}_{12} & 0 & {m}_{12}   \nonumber \\
\end{array} \right)}  
{\left\{ \begin{array}{cccc}
\bar{j}_{1} & {j}_{1} & \frac{1}{2}  \nonumber \\
\bar{j}_{2} & {j}_{2} & \frac{1}{2}  \nonumber \\
\bar{j}_{12} & {j}_{12} & 0  \nonumber \\
\end{array} \right \}} 
\langle \bar{j}_1||a^{\dagger}[1]||j_1\rangle 
\langle \bar{j}_2||a^{\dagger}[2]||j_2\rangle ~~~~~
\label{a1} 
\eea
In (\ref{a1}), 
$\left(a^{\dagger}[1] \otimes a^{\dagger}[2]\right)^0_0 \equiv 
\sum_{m,\bar{m}=\pm{\frac{1}{2}}} C^{0,0}_{\frac{1}{2},m;\frac{1}{2},\bar{m}} 
a^{\dagger}_m b^{\dagger}_{\bar{m}}$ with $a_{+\frac{1}{2}} \equiv a_1, 
~a_{-\frac{1}{2}} \equiv a_2$. 
The reduced matrix elements of the prepotential operators are  
given by:
\bea 
\langle \bar{j}||a||j\rangle  = \Pi({j,\bar{j}})~\delta_{\bar{j},j-\frac{1}{2}}, ~~~~    
\langle \bar{j}||a^{\dagger}||j\rangle  = \Pi({j,\bar{j}})~\delta_{\bar{j},j+\frac{1}{2}}.    
\label{rme}
\eea
The coefficients  ${\left( \begin{array}{cccc}
{j}_{1} & j_2 & {j}_{3}   \nonumber \\
{m}_{1} & m_2 & {m}_{3}   \nonumber \\
\end{array} \right)}$,   
${\left\{ \begin{array}{cccc}
{j}_{1} & {j}_{2} & j_{12}  \nonumber \\
{j}_{3} & {j}_{4} & j_{34}  \nonumber \\
{j}_{13} & {j}_{24} & j  \nonumber \\
\end{array} \right \}}$  represent $3j$ and $9j$ symbols 
\cite{varsha} respectively. 
Using the values: 
\bea 
{\left( \begin{array}{cccc}
\bar{j}_{12} & 0 & {j}_{12}   \nonumber \\
- \bar{m}_{12} & 0 & {m}_{12}   \nonumber \\ 
\end{array} \right)}   
& = & 
\frac{(-1)^{-\bar{j}_{12}+\bar{m}_{12}}}{\Pi(j_{12})}  
~~\delta_{j_{12},\bar{j}_{12}} 
\delta_{m_{12},\bar{m}_{12}},  \nonumber \\
{\left\{ \begin{array}{cccc}
\bar{j}_{1} & {j}_{1} & \frac{1}{2}  \nonumber \\
\bar{j}_{2} & {j}_{2} & \frac{1}{2}  \nonumber \\
\bar{j}_{12} & {j}_{12} & 0  \nonumber \\
\end{array} \right \}}  
& = & 
\frac{(-1)^{j_{1}+\frac{1}{2}+\bar{j}_{2}+\bar{j}_{12}}}{\Pi({\frac{1}{2},j_{12}})} 
~~ \delta_{j_{12},\bar{j}_{12}} 
{\left \{ \begin{array}{cccc}
\bar{j}_{1} &  {j}_{1} & \frac{1}{2}   \nonumber \\
j_{2} & \bar{j}_{2}  & {j}_{12}   \nonumber \\ 
\end{array} \right\}},   
\eea 
we get: 
\bea 
\langle \bar{j}_1,\bar{j}_2,\bar{j}_{12}, \bar{m}_{12}|
L_{12} |j_1,j_2,j_{12},m_{12} \rangle  
& = & \delta_{j_{12},\bar{j}_{12}} 
\delta_{m_{12},\bar{m}_{12}} 
(-1)^{j_{12}}  
\eta(j_1,\bar{j}_2) 
{\left \{ \begin{array}{cccc}
\bar{j}_{1} &  {j}_{1} & \frac{1}{2}   \nonumber \\
j_{2} & \bar{j}_{2}  & {j}_{12}   \nonumber \\ 
\end{array} \right\}}   
\nonumber \\ 
&& \langle \bar{j}_1||a^{\dagger}[1]||j_1\rangle 
\langle \bar{j}_2||a^{\dagger}[2]||j_2\rangle. 
\label{aa2} 
\eea

\noindent In (\ref{aa2}), the phase factor $\eta(a,b) \equiv (-1)^{a+b+\frac{1}{2}}$.  
Note that one can also  apply the above intertwining operator directly on the loop basis 
(\ref{fr}) to get it's matrix elements (\ref{aa2}) algebraically.   
Infact, this has provided an 
independent check on  the ``master formula" (\ref{aa2}) which is crucial for the 
computations below.  It simply states 
that  the intertwining operator $L_{12} = a^{\dagger}[1] \cdot \tilde{a}^{\dagger}[2]$ 
increases $j_1$ and $j_2$ by $\frac{1}{2}$. Further, as $L_{12}$ commutes with 
the $\left(J[1]+J[2]\right)^{2}$, the matrix elements are diagonal in 
$j_{12}$ and $m_{12}$.  The matrix elements of the intertwining operators 
in the geometrical form (\ref{aa2}) also tell us that 
all the 16 terms in $\sum_{\alpha\beta\gamma\delta=\pm} 
H^{}_{\alpha\beta\gamma\delta}$  in (\ref{abcd}) 
differ only in their  reduced matrix element structures. 
Therefore, we need to compute the matrix elements of only a single 
term in (\ref{abcd}), providing enormous simplification at the algebraic 
level.  In the following calculations, we choose this term to be the first 
term in (\ref{abcd}) associated with  the plaquette $abcd$ in 
Figure (\ref{fig:ABCD}):  
\bea
H^{}_{++++} \equiv  F_{abcd} \left(a^{\dagger}[1] \cdot \tilde{a}^{\dagger}[2]\right)_{a} 
\left (a^{\dagger}[2] \cdot\tilde{a}^{\dagger}[3]\right)_{b} 
\left (a^{\dagger}[3]\cdot\tilde{a}^{\dagger}[4]\right)_{c} 
\left (a^{\dagger}[4]\cdot\tilde{a}^{\dagger}[1]\right)_{d}F_{abcd} 
\label{ih1} 
\eea
As mentioned in the beginning of this section, we will work in $d=2$. The results are 
then easily generalized to arbitrary dimensions d and are given in the equation (\ref{dyna3}). 
In computing the loop dynamics below, it will be often convenient to change the angular 
momentum addition scheme\footnote{The relation (\ref{(12)(34)}) is proved by writing: 
\bea 
&& |j_1,j_2,j_3,j_4,j_{12}\rangle  \equiv   \sum_{all ~m} 
C^{~~0,0}_{j_{123}m_{123},j_4m_4}   
C^{j_{123},m_{123}}_{j_{12}m_{12},j_3m_3} 
|j_1,j_2,j_{12},m_{12}\rangle |j_{3}m_{3}\rangle |j_{4}m_{4}\rangle   
\nonumber \\
&=& \sum_{all ~m} \sum_{j_{34}} C^{0,0}_{j_{12}m_{12},j_{34}m_{34}}
|j_1,j_2,j_{12},m_{12}\rangle 
|j_3,j_4,j_{34},m_{34} \rangle = |(j_1,j_2)j_{12},(j_3,j_4)j_{12} \rangle.  \nonumber 
\eea
We  have used: 
\bea 
C^{~~0,0}_{j^a_{123}m^a_{123},j^a_4m^a_4}  =  \frac{(-1)^{j^a_4+m^a_4}}{\Pi(j^a_4)} 
\delta_{j^a_{123},j^a_4}\delta_{m^a_{123},-m^a_4}, ~~ 
\frac{(-1)^{j^a_4+m^a_4}}{\Pi(j^a_4)} 
C^{j^a_4-m^a_4}_{j^a_{12}m^a_{12},j^a_3m^a_3}  =  C^{0,0}_{j^a_{12}m^a_{12},j^a_{12}-m^a_{12}} 
C^{j^a_{12}-m^a_{12}}_{j^a_3m^a_3,j^a_4m^a_4}.  
\nonumber 
\eea \\ \\}: 
\bea 
|j_1,j_2,j_3,j_4,j_{12}\rangle  \equiv   
|j_1,j_2,j_{12},j_3,j_{123}(=j_{4}), j_{4},j_{total}=j_{(123)(4)}=0\rangle \nonumber \\ 
 =  |(j_1,j_2)j_{12},(j_3,j_4)j_{34}(=j_{12}), j_{total}=j_{(12)(34)} =0 \rangle
\equiv |(j_1,j_2)j_{12},(j_3,j_4)j_{12} \rangle
\label{(12)(34)} 
\eea
The equivalent scheme on the right of (\ref{(12)(34)}) simplifies the algebra. We also 
note that the 
normalization operator $F_{abcd}$ in (\ref{abcd}) has simple action on the loop states 
$|j_{abcd} \rangle$ defined in (\ref{plst})and (\ref{u1gl}): 
\bea 
F_{abcd}|j_{abcd} \rangle  = \frac{1}{\Pi({j_1,j_2,j_3,j_4})} |j_{abcd} \rangle    
\label{iof} 
\eea
Therefore, we only need to compute the matrix elements  of the 
intertwining operators in (\ref{ih1}) in the orthonormal loop 
basis given in (\ref{fr}).  

\noindent {\bf Loop dynamics at {\bf a}:}  

\vspace{0.4cm} 

In $H^{}_{++++}$ above,  the intertwining operator at a is $\left(a^{\dagger}[1] \cdot 
\tilde{a}^{\dagger}[2]\right)_{a}$. Using (\ref{aa2}), one directly gets: 
\bea 
\langle \bar{j}^a_1,\bar{j}^a_2,\bar{j}^a_3,\bar{j}^a_4,\bar{j}^a_{12}|
\left(a^{\dagger}[1] \cdot\tilde{a}^{\dagger}[2]\right)_{a}
|j^a_1,j^a_2,j^a_3,j^a_4,j^a_{12}\rangle  
 =  (-1)^{j^a_{12}}  \eta(j^a_1,\bar{j}^a_2) \delta_{j^a_{3},\bar{j}^a_{3}} 
\delta_{j^a_{4},\bar{j}^a_{4}} \delta_{j^a_{12},\bar{j}^a_{12}} 
\nonumber \\  
{\left \{ \begin{array}{cccc}
{j}^a_{1} &  \bar{j}^a_{1} & \frac{1}{2}  \\
\bar{j}^a_{2} & {j}^a_{2}  & {j}^a_{12}   \\ 
\end{array} \right\}}   
{\langle \bar{j}^a_1||a^{\dagger}[1]||
j^a_1\rangle \langle \bar{j}^a_2||a^{\dagger}[2]||j^a_2
\rangle}
\label{aa3} 
\eea
The intertwining operator $\left(a^{\dagger}[1] \cdot\tilde{a}^{\dagger}[2]\right)_{a}$ 
increases the SU(2) flux on the links $l_1$ as well as $l_4$ of Figure (\ref{fig:ABCD}). 
Note that this information is contained only in the last two reduced matrix element terms 
in (\ref{aa3}).  

\vspace{0.4cm} 

\noindent {\bf Loop dynamics at {\bf b}:}  

\vspace{0.4cm} 

The intertwining operator at b in (\ref{ih1}) is $\left(a^{\dagger}[2] \cdot 
\tilde{a}^{\dagger}[3]\right)_{b}$.  To compute it's action at b, we 
write the loop states (\ref{std2}) in terms of the basis states which 
diagonalize $(J[2]+J[3])^{2}$: 
\bea 
|j^{b}_1j^{b}_2j^{b}_3j^{b}_4j^{b}_{12} \rangle & = &
(-1)^{(j^{b}_1+j^{b}_2+j^{b}_3+j^{b}_{123})}\sum_{j^{b}_{23}} 
\Pi(j_{12}^{b},{j}_{23}^b) 
{\left\{ \begin{array}{cccc}
j^{b}_{1} & {j}^{b}_{2} & j^{b}_{12}  \\
{j}^{b}_{3} & j^{b}_{123} & j^{b}_{23}\\
\end{array} \right \}} |j^{b}_1j^{b}_2j^{b}_3j^{b}_4j^{b}_{23} \rangle \nonumber \\
&=& (-1)^{(j^b_1+j^b_2+j^b_3+j^b_4)} \sum_{j^{b}_{23}} 
\Pi(j^{b}_{12}, j^{b}_{23}) 
{\left\{ \begin{array}{cccc}
j^{b}_{1} & {j}^{b}_{2} & j^{b}_{12}  \\
{j}^{b}_{3} & j^{b}_{4} & j^{b}_{23}\\
\end{array} \right \}} |j^{b}_1j^{b}_2j^{b}_3j^{b}_4j^{b}_{23} \rangle 
\label{a4} 
\eea
In (\ref{a4}), $|j^{b}_1j^{b}_2j^{b}_3j^{b}_4j^{b}_{23} \rangle 
\equiv |j^{b}_1,(j^{b}_2j^{b}_3),j^b_{23},j^b_{123},j^b_4,j^b_{1234}
=m^b_{1234}=0\rangle$ and 
we have used $j_{123}=j_{4}$. Note that the phase factor 
$(-1)^{(j^b_1+j^b_2+j^b_3+j^b_4)}$ in (\ref{a4}) is real because of 
the triangular constraints on the angular momenta or equivalently 
(\ref{part}). Now we use:  
\bea 
|j^{b}_1j^{b}_2j^{b}_3j^{b}_4j^{b}_{23} \rangle = 
(-1)^{2j^{b}_{1}} |j^{b}_2j^{b}_3j^{b}_4j^{b}_1j^{b}_{23}=j_{41}^b \rangle
\label{j23j41} 
\eea
and (\ref{aa2}) to get: 
\bea 
\langle \bar{j}^b_1,\bar{j}^b_2,\bar{j}^b_3,\bar{j}^b_4,\bar{j}^b_{12}|
\left(a^{\dagger}[2] \cdot\tilde{a}^{\dagger}[3]\right)_{b}
|j^b_1,j^b_2,j^b_3,j^b_4,j^b_{12}\rangle    =  
(-1)^{j_2^b+j_3^b-\bar{j}_2^b-\bar{j}_3^b} \eta(j_2^b,\bar{j}_3^b) 
\delta_{j_1^b,\bar{j}_1^b} \delta_{j_4^b,\bar{j}_4^b}  
\Pi(j^{b}_{12},\bar{j}^{b}_{12}) 
 \nonumber \\ \nonumber \\   
\sum_{j^{b}_{23}} (-)^{j_{23}^b} 
\Pi(j^{b}_{23}) 
{\left\{ \begin{array}{cccc}
j^{b}_{3} & {j}^{b}_{2} & j^{b}_{23}  \\
{j}^{b}_{1} & j^{b}_{4} & j^{b}_{12}\\
\end{array} \right \}}  
{\left\{ \begin{array}{cccc}
j^{b}_{1} & \bar{j}^{b}_{4}  & {j}^{b}_{23} \\
\bar{j}^{b}_{3}  & j^{b}_{2} & \bar{j}^{b}_{12}\\
\end{array} \right \}} 
{\left\{ \begin{array}{cccc}
\bar{j}_{3}^b & \bar{j}_{2}^b & j_{23}^b  \\
{j}_{2}^b & {j}_{3}^b & \frac{1}{2}\\
\end{array} \right \}} 
 \langle \bar{j}^b_2||a^{\dagger}[2]||j^b_2\rangle 
\langle \bar{j}^b_3||a^{\dagger}[3]||j^b_3\rangle \nonumber  
\eea
The summation over $j_{23}^b$ in the last line above  can be performed using 
Biedenharn-Elliot identity \cite{varsha}: 
\bea 
\sum_{x} (-1)^{x} \Pi^2(x) 
{\left\{ \begin{array}{cccc}
a & b  & x  \\
c & d & p \\
\end{array} \right \}}  
{\left\{ \begin{array}{cccc}
c  & d  & x  \\
e & f  & q \\
\end{array} \right \}}  
{\left\{ \begin{array}{cccc}
e & f &  x \\
b & a & s \\
\end{array} \right \}} = 
(-1)^{-r} {\left\{ \begin{array}{cccc}
p & q &  s \\
e & a  & d \\
\end{array} \right \}}  \nonumber 
{\left\{ \begin{array}{cccc}
p & q & s  \\
f & b & c \\
\end{array} \right \}} \nonumber \\
r = (a+b+c+d+e+f+p+q+s).  \nonumber   
\eea 
Finally, the loop dynamics at lattice site b is given by: 
\bea 
\langle \bar{j}^b_1,\bar{j}^b_2,\bar{j}^b_3,\bar{j}^b_4,\bar{j}^b_{12}|
\left(a^{\dagger}[2] \cdot\tilde{a}^{\dagger}[3]\right)_{b}
|j^b_1,j^b_2,j^b_3,j^b_4,j^b_{12}\rangle =   
(-1)^{j_1^b+j_4^b} \eta(j_2^b,\bar{j}_3^b)  \eta(j_{12}^b,\bar{j}_{12}^b)
\delta_{j_1^b,\bar{j}_1^b} \delta_{j_4^b,\bar{j}_4^b}  
 \nonumber \\ \nonumber \\   
\Pi(j^{b}_{12}, \bar{j}^{b}_{12})
{\left\{ \begin{array}{cccc}
{j}^b_{12} & \bar{j}^b_{12} & \frac{1}{2}  \\
\bar{j}^b_{3} & j^b_{3} & j_4^b\\
\end{array} \right \}}  
{\left\{ \begin{array}{cccc}
{j}^b_{12} & \bar{j}^b_{12} & \frac{1}{2}  \\
\bar{j}^b_{2} & j^b_{2} & j_1^b\\
\end{array} \right \}} 
 \langle \bar{j}^b_2||a^{\dagger}[2]||j^b_2\rangle 
\langle \bar{j}^b_3||a^{\dagger}[3]||j^b_3\rangle 
\label{meb+}
\eea

\noindent {\bf Loop dynamics at {\bf c}:}  

\vspace{0.4cm} 

\noindent At c, we use (\ref{(12)(34)}) to write 
\bea 
|j^c_1,j^c_2,j^c_3,j^c_4,j^c_{12}\rangle =  |(j^c_1,j^c_2)j_{12}^c,(j^c_3j^c_4),j^c_{12}\rangle
= (-1)^{2j_{12}^c}|(j_3^c,j_4^c)j_{12}^c,(j_1^cj_2^c)j_{12}^c\rangle \nonumber 
\eea
to get: 
\bea 
\langle \bar{j}^c_1,\bar{j}^c_2,\bar{j}^c_3,\bar{j}^c_4,\bar{j}^c_{12}|
\left(a^{\dagger}[3] \cdot\tilde{a}^{\dagger}[4]\right)_{c}
|j^c_1,j^c_2,j^c_3,j^c_4,j^c_{12}\rangle   
 =  (-1)^{j^c_{12}}  \eta(j^c_3,\bar{j}^c_4) \delta_{j^c_{1},\bar{j}^c_{1}} 
\delta_{j^c_{2},\bar{j}^c_{2}} \delta_{j^c_{12},\bar{j}^c_{12}} 
\nonumber  \\ 
{\left \{ \begin{array}{cccc}
{j}^c_{3} &  \bar{j}^c_{3} & \frac{1}{2}  \\
\bar{j}^c_{4} & {j}^c_{4}  & {j}^c_{12}   \\ 
\end{array} \right\}}   
 \langle \bar{j}^c_3||a^{\dagger}[3]||j^c_3\rangle 
\langle \bar{j}^c_4||a^{\dagger}[4]||j^c_4\rangle 
\label{aa4} 
\eea

\noindent {\bf Loop dynamics at {\bf d}:} 

\vspace{0.4cm} 

\noindent To compute the loop dynamics at  d, we write: 
\bea 
&& \hspace{2cm} |j_1^d,j_2^d,j_3^d,j_4^d,j_{12}^d\rangle  =  
|j_1^d,j_2^d,j_{12}^d,j_3^d,j_4^d,j_{34}^d(=j_{12}^d)\rangle   
\nonumber \\
& = & \sum_{j^d_{14}} (-1)^{j_3^d+j_4^d-j_{12}^d} 
\Pi(j^d_{12}, j^d_{34}, j^d_{14}, j^d_{23}) 
{\left\{ \begin{array}{cccc}
{j}^d_{1} & {j}^d_{2} & j^d_{12}  \nonumber \\
{j}^d_{4} & {j}^d_{3} & j^d_{34}=j^d_{12}   \nonumber \\
{j}^d_{14} & {j}^d_{23} = j^d_{14} & 0  \nonumber \\
\end{array} \right \}} 
|j_4^d,j_1^d,j_{14}^d,j_2^d,j_3^d,j_{23}^d\rangle  \nonumber \\
& = & (-1)^{2j_4^d} 
\sum_{j_{14}} 
\Pi(j^d_{12}, j^d_{14}) 
(-1)^{j_1^d+j_2^d+j_3^d+j_4^d} 
{\left\{ \begin{array}{cccc}
{j}^d_{1} & {j}^d_{2} & j^d_{12}   \\
{j}^d_{3} & {j}^d_{3} & j^d_{14}   \\
\end{array} \right \}} 
|j_4^d,j_1^d,j_{14}^d,j_2^d,j_3^d,j_{23}^d\rangle  
\label{xyz} 
\eea 
In (\ref{xyz}), we have used  
${\left\{ \begin{array}{cccc}
{j}^d_{1} & {j}^d_{2} & j^d_{12}  \nonumber \\
{j}^d_{4} & {j}^d_{3} & j^d_{12}   \nonumber \\
{j}^d_{14} & j^d_{14} & 0  \nonumber \\
\end{array} \right \}} = {(-1)^{j_2^d+j_{12}^d+j_4^d+j_{14}^d}}
\left(\Pi (j_{12}^d, j_{14}^d)\right)^{-1}  
{\left\{ \begin{array}{cccc}
{j}^d_{1} & {j}^d_{2} & j^d_{12}  \nonumber \\
{j}^d_{3} & {j}^d_{3} & j^d_{14}   \nonumber \\
\end{array} \right \}}$. \\  \\
Finally, using (\ref{xyz}) and the Biedenharn-Elliot identity, the dynamics at d is given by: 
\bea 
\langle \bar{j}^d_1,\bar{j}^d_2,\bar{j}^d_3,\bar{j}^d_4,\bar{j}^d_{12}|
\left(a^{\dagger}[4] \cdot\tilde{a}^{\dagger}[1]\right)_{d}
|j^d_1,j^d_2,j^d_3,j^d_4,j^d_{12}\rangle =   
- (-1)^{j_2^d+j_3^d} \eta(j_4^d,\bar{j}_1^d)  \eta(j_{12}^d,\bar{j}_{12}^d)
\delta_{j_2^d,\bar{j}_2^d} \delta_{j_3^d,\bar{j}_3^d}  
 \nonumber \\ \nonumber \\   
\Pi(j^{d}_{12}, \bar{j}^{d}_{12}) 
{\left\{ \begin{array}{cccc}
{j}^d_{12} & \bar{j}^d_{12} & \frac{1}{2}  \\
\bar{j}^d_{1} & j^d_{1} & j_2^d\\
\end{array} \right \}}  
{\left\{ \begin{array}{cccc}
{j}^d_{12} & \bar{j}^d_{12} & \frac{1}{2}  \\
\bar{j}^d_{4} & j^d_{4} & j_3^d\\
\end{array} \right \}} 
 \langle \bar{j}^d_4||a^{\dagger}[4]||j^d_4\rangle 
 \langle \bar{j}^d_1||a^{\dagger}[1]||j^d_1\rangle 
\label{meb++}
\eea 

\vspace{0.4cm} 

\noindent{\bf Loop dynamics at {\bf abcd}:} 

\vspace{0.4cm} 

\noindent We now weave or glue the dynamics at a,b,c,d with the help of U(1) Gauss law 
(\ref{u1gl}):  
$j^{a}_{1} = j^{b}_{3} \equiv j_1, ~j^{b}_{2} = j^{c}_{4} \equiv j_2,~  
j^{c}_{3} = j^{d}_{1} \equiv j_3, ~ j^{d}_{4} = j^{a}_{2} \equiv j_4$ and 
$\bar{j}^{a}_{1} = \bar{j}^{b}_{3} \equiv 
\bar{j}_1, ~\bar{j}^{b}_{2} = \bar{j}^{c}_{4} \equiv \bar{j}_2,~  
\bar{j}^{c}_{3} = \bar{j}^{d}_{1} \equiv \bar{j}_3, ~ 
\bar{j}^{d}_{4} = \bar{j}^{a}_{2} \equiv \bar{j}_4$.  
This implies:  
\bea 
\langle \bar{j}^a_1||a^{\dagger}[1]|| j^a_1\rangle  =  
\langle \bar{j}^b_3||a^{\dagger}[3]||j^b_3 \rangle   =    
\Pi({j_1,\bar{j}_1}) \delta_{\bar{j}_1,j_1+\frac{1}{2}},~   
\langle \bar{j}^b_2||a^{\dagger}[2]|| j^b_2\rangle   = 
\langle \bar{j}^c_4||a^{\dagger}[4]||j^c_4 \rangle    =   
\Pi({j_2,\bar{j}_2})\delta_{\bar{j}_2,j_2+\frac{1}{2}} 
\nonumber \\
\langle \bar{j}^c_3||a^{\dagger}[3]|| j^c_3\rangle =  
\langle \bar{j}^d_1||a^{\dagger}[1]||j^d_1 \rangle    =   
\Pi({j_3,\bar{j}_3}) \delta_{\bar{j}_3,j_3+\frac{1}{2}},~  
\langle \bar{j}^d_4||a^{\dagger}[4]|| j^d_4\rangle = \langle \bar{j}^a_2||a^{\dagger}[2]||j^a_2 \rangle 
 = \Pi({j_4,\bar{j}_4})\delta_{\bar{j}_4,j_4+\frac{1}{2}} \nonumber  
\eea \\
Using the U(1) identifications and  (\ref{iof}), (\ref{aa3}), 
(\ref{meb+}), (\ref{aa4}) and (\ref{meb++}) and merging all these 
equations carefully,  we get: 
\bea 
&& \langle \bar{j}_{abcd}|{\textrm Tr}U_{abcd}| {j}_{abcd} \rangle =  
- \delta_{j^a_{3},\bar{j}^a_{3}} 
\delta_{j^a_{4},\bar{j}^a_{4}} 
\delta_{j^a_{12},\bar{j}^a_{12}}  
\delta_{j_1^b,\bar{j}_1^b} \delta_{j_4^b,\bar{j}_4^b}  
\delta_{j^c_{1},\bar{j}^c_{1}} \delta_{j^c_{2},\bar{j}^c_{2}} \delta_{j^c_{12},\bar{j}^c_{12}} 
\delta_{j_2^d,\bar{j}_2^d} \delta_{j_3^d,\bar{j}_3^d}  
(-1)^{j^a_{12}+j_{12}^c}
\nonumber \\
\nonumber \\
&&(-1)^{j_1^b+j_4^b+j_2^d+j_3^d}  
\bar{\Pi}({{j_1}\bar{j_1}}) 
\bar{\Pi}({{j_2}\bar{j_2}}) 
\bar{\Pi}({{j_3}\bar{j_3}}) 
\bar{\Pi}({{j_4}\bar{j_4}}) 
\bar{\Pi}({{j^b_{12}}\bar{j^b_{12}}}) 
\bar{\Pi}({{j^d_{12}}\bar{j^d_{12}}}) 
{\left \{ \begin{array}{cccc}
{j}_{1} &  \bar{j}_{1} & \frac{1}{2}  \\
\bar{j}_{4} & {j}_{4}  & {j}^a_{12}   \\ 
\end{array} \right\}}   
{\left\{ \begin{array}{cccc}
{j}^b_{12} & \bar{j}^b_{12} & \frac{1}{2}  \\
\bar{j}_{1} & j_{1} & j_4^b\\
\end{array} \right \}}  
\nonumber  \\ 
\nonumber  \\ 
&&
{\left\{ \begin{array}{cccc}
{j}^b_{12} & \bar{j}^b_{12} & \frac{1}{2}  \\
\bar{j}_{2} & j_{2} & j_1^b\\
\end{array} \right \}} 
{\left \{ \begin{array}{cccc}
{j}_{3} &  \bar{j}_{3} & \frac{1}{2}  \\
\bar{j}_{2} & {j}_{2}  & {j}^c_{12}   \\ 
\end{array} \right\}}   
{\left\{ \begin{array}{cccc}
{j}^d_{12} & \bar{j}^d_{12} & \frac{1}{2}  \\
\bar{j}_{3} & j_{3} & j_2^d\\
\end{array} \right \}}  
{\left\{ \begin{array}{cccc}
{j}^d_{12} & \bar{j}^d_{12} & \frac{1}{2}  \\
\bar{j}_{4} & j_{4} & j_3^d \\
\end{array}\right\}} 
\label{diff} 
\eea
In (\ref{diff}), $\bar{\Pi}(a,b) \equiv (-1)^{a+b+\frac{1}{2}}\Pi(a,b)$.
Note that $\bar{\Pi}(a,b)$  are symmetric $\bar{\Pi}(a,b)= \bar{\Pi}(b,a)$  and real. 
We have ignored the 16 $\delta$ functions 
$\prod_{i=1}^{4} \left(\delta_{\bar{j}_{i},j_{i} +\frac{1}{2}} + 
\delta_{\bar{j}_{i},j_{i} -\frac{1}{2}}\right)$   
coming from the reduced matrix elements in (\ref{rme}) 
as they are already contained in the six $6j$ symbols in (\ref{diff}).   
The above d=2 loop dynamics  and it's generalization to arbitrary 
dimensions are discussed in detail in sections (3.1) and (3.2).

\end{document}